\documentclass[prx,twocolumn,superscriptaddress,amsmath,amsthm,showpacs]{revtex4}
\usepackage{amssymb}
\usepackage{color}
\usepackage{graphicx}
\usepackage{braket}

\newcommand{\ketbra}[2]{| #1 \rangle \langle #2 |}
\newcommand{\expect}[1]{\langle #1 \rangle}

\begin{document}
\title{Revealing nonclassicality beyond Gaussian states via a single marginal distribution }
\author{Jiyong Park}
\affiliation{Department of Physics, Texas A\&M University at Qatar, Education City, P.O.Box 23874, Doha, Qatar}
\author{Yao Lu}
\affiliation{Center for Quantum Information, Institute for Interdisciplinary Information Sciences, Tsinghua University, Beijing 100084, People’s Republic of China}
\author{Jaehak Lee}
\affiliation{Department of Physics, Texas A\&M University at Qatar, Education City, P.O.Box 23874, Doha, Qatar}
\author{Yangchao Shen}
\affiliation{Center for Quantum Information, Institute for Interdisciplinary Information Sciences, Tsinghua University, Beijing 100084, People’s Republic of China}
\author{Kuan Zhang}
\affiliation{Center for Quantum Information, Institute for Interdisciplinary Information Sciences, Tsinghua University, Beijing 100084, People’s Republic of China} 
\author{Shuaining Zhang}
\affiliation{Center for Quantum Information, Institute for Interdisciplinary Information Sciences, Tsinghua University, Beijing 100084, People’s Republic of China}
\author{M. Suhail Zubairy}
\affiliation{Department of Physics and Institute of Quantum Studies, Texas A\&M University, College Station, TX 77843, USA}
\author{Kihwan Kim}
\affiliation{Center for Quantum Information, Institute for Interdisciplinary Information Sciences, Tsinghua University, Beijing 100084, People’s Republic of China}
\author{Hyunchul Nha}
\affiliation{Department of Physics, Texas A\&M University at Qatar, Education City, P.O.Box 23874, Doha, Qatar}
\affiliation{E-mail: hyunchul.nha@qatar.tamu.edu}
\date{\today}


\begin{abstract}
A standard method to obtain information on a quantum state is to measure marginal distributions along many different axes in phase space, which forms a basis of quantum state tomography. We theoretically propose and experimentally demonstrate a general framework to manifest nonclassicality by observing a single marginal distribution only, which provides a novel insight into nonclassicality and a practical applicability to various quantum systems. Our approach maps the 1-dim marginal distribution into a factorized 2-dim distribution by multiplying the measured distribution or the vacuum-state distribution along an orthogonal axis. The resulting fictitious Wigner function becomes unphysical only for a nonclassical state, thus the negativity of the corresponding density operator provides an evidence of nonclassicality. Furthermore, the negativity measured this way yields a lower bound for entanglement potential---a measure of entanglement generated using a nonclassical state with a beam splitter setting that is a prototypical model to produce continuous-variable (CV) entangled states. Our approach detects both Gaussian and non-Gaussian nonclassical states in a reliable and efficient manner. Remarkably, it works regardless of measurement axis for all non-Gaussian states in finite-dimensional Fock space of any size, also extending to infinite-dimensional states of experimental relevance for CV quantum informatics. We experimentally illustrate the power of our criterion for motional states of a trapped ion confirming their nonclassicality in a measurement-axis independent manner. We also address an extension of our approach combined with phase-shift operations, which leads to a stronger test of nonclassicality, i.e. detection of genuine non-Gaussianity under a CV measurement.
\end{abstract}

\maketitle

Nonclassicality is a fundamentally profound concept to identify quantum phenomena inaccessible from classical physics. It also provides a practically useful resource, e.g. entanglement, making possible a lot of applications in quantum information processing beyond classical counterparts \cite{ReviewCVQI, Cerf, ReviewGQI}. 
A wide range of quantum systems, e.g. field amplitudes of light, collective spins of atomic ensembles, and motional modes of trapped ions, Bose-Einstein condensate and mechanical oscillators, can be employed for quantum information processing based on continuous variables (CVs) \cite{Cerf}. It is of crucial importance to establish efficient and reliable criteria of nonclassicality for CV systems, desirably testable with less experimental resources, e.g. fewer measurement settings \cite{Richter2002,Richter2002', Mari2011, Park2015, ParkNha} and with the capability of detecting a broad class of nonclassical states. In this paper, in view of the Glauber-Sudarshan P-function \cite{Glauber1963,Sudarshan1963}, those states that cannot be represented as a convex mixture of coherent states are referred to as nonclassical.

A standard method to obtain information on a CV quantum state is to measure marginal distributions along many different axes in phase space constituting quantum state tomography \cite{Lvovsky2009}. 
This tomographic reconstruction may reveal nonclassicality to some extent, e.g. negativity of Wigner function making only a subset of whole nonclassicality conditions. However it typically suffers from a legitimacy problem, i.e., the measured distributions do not yield a physical state when directly employed due to finite data and finite binning size \cite{D'Ariano1994, Lvovsky2009}. Much efforts was made to employ estimation methods finding a most probable quantum state closest to the obtained data \cite{Hradil1997, Banaszek1999, Rehacek2010, Teo2011}. There were also numerous studies to directly detect nonclassicality, e.g. an increasingly large number of hierarchical conditions \cite{Richter2002} requiring information on two or more marginal distributions or measurement of many higher-order moments \cite{Vogel,Vogel',Bednorz}. An exception would be the case of Gaussian states, with its nonclassical squeezing demonstrated by the variance of distribution along a squeezed axis.

Here we theoretically propose and experimentally demonstrate a simple, powerful, method to directly manifest nonclassicality by observing a {\it single} marginal distribution applicable to a wide range of nonclassical states. 
Our approach makes use of a phase-space map that transforms the marginal distribution (obtained from measurement) to a factorized Wigner distribution by multiplying the same distribution or the vacuum-state distribution along an orthogonal axis. We refer to those mathematical procedures as demarginalization maps (DMs), since a 1-dimensional marginal distribution is converted to a fictitious 2-dimensional Wigner function. The same method can be applied equally to the characteristic function as well as the Wigner function. We show that a classical state, i.e. a mixture of coherent states, must yield a physical state under our DMs. That is, the unphysicality emerging under DMs is a clear signature of nonclassicality. Remarkably, for all non-Gaussian states in finite dimensional space \cite{PN}, our test works for an arbitrary single marginal distribution thus experimentally favorable. It also extends to non-Gaussian states in infinite dimension, particularly those without squeezing effect. We introduce a quantitative measure of nonclassicality using our DMs, which provides a lower bound of entanglement potential \cite{Asboth2005}---an entanglement measure under a beam-splitter setting versatile for CV entanglement generation \cite{Kim2002, Wang, Asboth2005, Tahira2009}. Along this way, our method makes a rigorous connection between single-mode nonclassicality and negative partial transpose (NPT) entanglement \cite{Peres, Horodecki, Duan, Simon, MM, MM', Walborn, Walborn', Nha2008, Nha2012}, which bears on entanglement distillation \cite{EntanglementDistillation} and nonlocality \cite{PeresConjecture, PC1, PC2, PC3, PC4, PC5, PC6, PC7}. 

As the measurement of a marginal distribution is highly efficient in various quantum systems, e.g. homodyne detection in quantum optics, our proposed approach can provide a practically useful and reliable tool in a wide range of investigations for CV quantum physics.  
We here experimentally illustrate the power of our approach by manifesting nonclassicality of motional states in a trapped-ion system. Specifically, we confirm the nonclassicality regardless of measured quadrature axis by introducing a simple faithful test using only a subset of data points, not requiring data manipulation under numerical methods unlike the case of state reconstruction. We also extend our approach combined with phase-randomization in order to obtain a criterion on genuine non-Gaussianity.


\section{Demarginalization maps and nonclassicality measure}
\subsection{Nonclassicality test via demarginalization maps}
We first introduce our main tools, i.e. demarginalization maps (DMs),
	\begin{align}
		\mathcal{D}_{1} & : W_{\rho} ( q, p ) \mapsto M_{\rho} ( x ) M_{\rho} ( y )\equiv W_{\rho}^{\rm DM1} ( x, y ), \label{eq:DM1} \\
		\mathcal{D}_{2} & : W_{\rho} ( q, p ) \mapsto M_{\rho} ( x ) M_{\ketbra{0}{0}} ( y )\equiv W_{\rho}^{\rm DM2} ( x, y ), \label{eq:DM2}
	\end{align}
where $( x, y )^{T} = \mathcal{R} ( \theta ) ( q, p )^{T}$ is a pair of orthogonal quadratures rotated from position $q$ and momentum $p$ with $
		\mathcal{R} ( \theta ) =
		\begin{pmatrix}
			\cos \theta & \sin \theta \\
			- \sin \theta & \cos \theta
		\end{pmatrix}$. $M_{\rho} ( x ) = \int dy W_{\rho} ( x, y )$ is a marginal distribution of the Wigner function $W_{\rho} ( q, p ) = \frac{2}{\pi} \mathrm{tr} [ \rho \hat{D} ( \alpha ) (-1)^{\hat{n}}\hat{D}^{\dag} ( \alpha )  ]$, where $\hat{D} ( \alpha ) = e^{ \alpha \hat{a}^{\dag} - \alpha^{*} \hat{a}}$ is a displacement operator with $\alpha = q + ip$ \cite{Barnett, Scully}. 

Our DM methods proceed as follows. Given a state with its Wigner function $W_{\rho} ( q, p )$, we measure a marginal distribution $M_{\rho} ( x )$ along a certain axis, $x=q\cos\theta+p\sin\theta$. We then construct a fictitious, factorized, Wigner function $W_{\rho}^{\rm DM} ( x, y )$ either by replicating the obtained distribution as $M_{\rho} ( x ) M_{\rho} ( y )$ (DM1) or by multiplying the marginal distribution of a vacuum state as $M_{\rho} ( x ) M_{\ketbra{0}{0}} ( y )$  (DM2),
with $M_{\ketbra{0}{0}} ( y )=\sqrt{\frac{2}{\pi}}e^{-2y^2}$ (Fig. 1.). We test whether $W_{\rho}^{\rm DM} ( x, y )$ is a legitimate Wigner function to represent a physical state.

{\bf Nonclassicality criteria}---The constructed functions in Eqs. \ref{eq:DM1} and \ref{eq:DM2} are both in factorized forms, so judging their legitimacy is related to the problem what quantum states can possess a factorized Wigner function \cite{Note1}. Every coherent state $|\beta\rangle$ has a factorized Wigner function against all pairs of orthogonal quadratures, $W_{\ketbra{\beta}{\beta}} ( x, y ) = \frac{2}{\pi} e^{- 2 ( x - \beta_{x} )^{2}}e^{ - 2( y - \beta_{y} )^{2}}$ \cite{Park2015}. 
Owing to this factorizability, the maps $\mathcal{D}_{1}$ and $\mathcal{D}_{2}$ transform a classical state into another classical one. 
A mixture of coherent states has a Wigner function 	
\begin{equation} \label{eq:WC1}
		W_{\rho_{\mathrm{cl}}} ( x, y ) = \int d^{2}\beta P ( \beta_{x}, \beta_{y} ) W_{\ketbra{\beta}{\beta}} ( x, y ),
	\end{equation}
with the probability density $P ( \beta_{x}, \beta_{y} )$ for a coherent state 
$|\beta\rangle$ ($\beta= \beta_{x} + i \beta_{y}$).
Applying each DM leads to
	\begin{align} \label{eq:WC2}
		\mathcal{D}_{j} [ W_{\rho_{\mathrm{cl}}} ( q, p ) ]= \int d^{2} \beta Q_{1} ( \beta_{x} ) Q_{j} ( \beta_{y} ) W_{\ketbra{\beta}{\beta}} ( x, y )
	\end{align}
$(j = 1,2)$, where $Q_{1} (\beta_{x}) = \int d\beta_{y} P ( \beta_{x}, \beta_{y} )$ and $Q_{2} (\beta_{x} ) = \delta ( \beta_{x} )$ are nonnegative. 
The resulting distributions in Eq.~\ref{eq:WC2} also represent a certain mixture of coherent states, hence a physical state. Therefore, if an unphysical Wigner function emerges under our DMs, the input state must be nonclassical.

{\bf Gaussian states}---Let us first consider a Gaussian state $\sigma$ that has a squeezed quadrature $\hat{x}$ with $V_{x} \equiv \Delta^2 \hat{x}< \frac{1}{4}$. Taking the squeezed marginal $M_{\sigma} ( x ) = \frac{1}{\sqrt{2 \pi V_{x}}} e^{- \frac{(x - \expect{\hat{x}})^{2}}{2V_{x}}}$ yields
	\begin{align}
		\mathcal{D}_{1} [ W_{\sigma} ( q, p ) ] & = \frac{1}{2 \pi V_{x}} e^{- \frac{(x-\expect{\hat{x}})^{2}}{2V_{x}}} e^{-\frac{(y-\expect{\hat{x}})^{2}}{2V_{x}}}, \nonumber \\
		\mathcal{D}_{2} [ W_{\sigma} ( q, p ) ] & = \frac{1}{\pi \sqrt{V_{x}}} e^{- \frac{(x-\expect{\hat{x}})^{2}}{2V_{x}}}e^{- 2y^{2}},
	\end{align}
both of which violate the uncertainty relation $\Delta \hat{x}\Delta \hat{y}\geq \frac{1}{4}$. Thus, the squeezed state turns into an unphysical state under our DMs. 
This method, of course, succeeds only when the observed marginal distribution is along a squeezed axis that generally extends to a finite range of angles, if not the whole angles \cite{supple}. 
We can further make the test successful {\it regardless of quadrature axis} by introducing a random phase rotation on a Gaussian state \cite{supple}. Note that a mixture of phase-rotations, which transforms a Gaussian to a non-Gaussian state, does not create nonclassicality, so the nonclassicality detected after phase rotations is attributed to that of the original state.

{\bf Non-Gaussian states}---More importantly, we now address non-Gaussian states. 
Every finite-dimensional state (FDS) in Fock basis, i.e. $\rho = \sum_{j,k = 0}^{N} \rho_{jk} \ketbra{j}{k}$ is nonclassical, since all coherent states (except vacuum), and their mixtures as well, have an extension to infinite Fock states. It is nontrivial to demonstrate the nonclassicality of FDS when one has access to limited information, e.g., a noisy state $f \ketbra{0}{0} + (1-f) \ketbra{1}{1}$ for $f\ge\frac{1}{2}$ has no simple signatures of nonclassicality like squeezing and negativity of Wigner function. 
We prove that our DMs are able to detect all non-Gaussian states in finite dimension of any size, with details in Sec. S4 of \cite{supple}. The essence of our proof is that there always exists a submatrix of the density operator corresponding to DMs, which is not positive-definite.
Remarkably, this non-positivity emerges for a marginal distribution along an arbitrary direction, which means that the nonclassicality of FDS is confirmed regardless of the quadrature axis measured, just like the phase-randomized Gaussian states introduced in \cite{supple}. This makes our DM test experimentally favorable, while the degree of negativity may well depend on the quadrature axis except rotationally-symmetric states. Our criteria can further be extended to non-Gaussian states in infinite dimension, particularly those without squeezing effect \cite{supple}. 

As an illustration, we show the case of a FDS $|\Psi\rangle=\frac{1}{\sqrt{2}}\left(|0\rangle+|2\rangle\right)$, whose original Wigner function and matrix elements are displayed in Fig. 1 (a). 
Our DM methods yield matrix elements as shown in Fig. 1 (b) and (c). The non-positivity of the density operator is then demonstrated by, e.g., $\langle 0|\rho|0\rangle\langle 8|\rho|8\rangle -|\langle 0|\rho|8\rangle|^2<0$ under DM1 and $\langle 0|\rho|0\rangle\langle 4|\rho|4\rangle -|\langle 0|\rho|4\rangle|^2<0$ under DM2, respectively. 


	\begin{figure}[!t]
	\centering
	\includegraphics[ scale = 0.35]{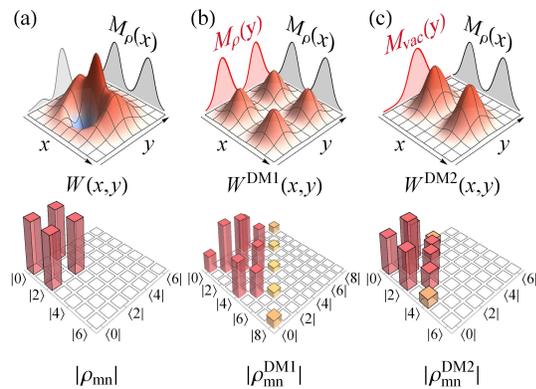}
		\caption{Illustrating DMs (a) original Wigner function $W_{\rho}(x,y)$ of $|\Psi\rangle=\frac{1}{\sqrt{2}}\left(|0\rangle+|2\rangle\right)$, with its marginal $M_{\rho}(x)=\int dyW_{\rho}(x,y)$ in the backdrop, (b) a fictitious Wigner function $W^{\rm DM1}(x,y)\equiv M_{\rho} ( x ) M_{\rho} ( y )$, with the same distribution $M_{\rho}(y)$ replicated along the orthogonal axis (red solid curve), and (c) $W^{\rm DM2}(x,y)\equiv M_{\rho} ( x ) M_{\ketbra{0}{0}} ( y )$, with the vacuum-state distribution $M_{\ketbra{0}{0}} ( y )$ used (red solid curve). The second panels show the corresponding density matrix elements. $W^{\rm DM1}(x,y)$ and $W^{\rm DM2}(x,y)$ in (b) and (c) do not represent any physical states, confirming the nonclassicality of $|\Psi\rangle$.}  
		\label{fig:DM}
	\end{figure}
\subsection{Nonclassicality measure and entanglement potential} 
We may define a measure of nonclassicality using our DMs as
	\begin{equation} \label{eq:DMN}
			\mathcal{N}_{\mathrm{DM}} ( \rho ) \equiv\max_{\theta \in ( 0, \pi )} \frac{|| \rho_{\mathrm{DM}}^{\theta} ||_{1} - 1}{2},
	\end{equation}
where $||\cdot||_{1}$ is a trace norm and $\rho_{\mathrm{DM}}^{\theta}$ a density matrix under DM using a marginal distribution at angle $\theta$.
Our DM negativity possesses the following properties appropriate as a nonclassicality measure, with details in \cite{supple}. 
(i) $\mathcal{N}_{\mathrm{DM}}= 0$ for a classical state, (ii) convex, i.e. non-increasing via mixing states, 
$\mathcal{N}_{\mathrm{DM}}( \sum_{j} p_{j} \rho_{j} ) \leq \sum_{j} p_{j} \mathcal{N}_{\mathrm{DM}} ( \rho_{j} )$, and (iii) invariant under a classicality-preserving unitary operation,
$\mathcal{N}_{\mathrm{DM}} ( \hat{U}_{c} \rho \hat{U}_{c}^{\dag} ) = \mathcal{N}_{\mathrm{DM}} ( \rho )$, where $\hat{U}_{c}$ refers to displacement or phase rotation. Combining (ii) and (iii), we also deduce the property that (iv) $\mathcal{N}_{\mathrm{DM}}$ does not increase under generic classicality preserving operations (mixture of unitary operations). 

Our nonclassicality measure also makes a significant connection to entanglement potential as follows.	
A prototypical scheme to generate a CV entangled state is to inject a single-mode nonclassical state into a beam splitter (BS) \cite{Kim2002, Wang, Asboth2005, Tahira2009}. It is important to know the property of those entangled states under PT, which bears on the distillibility of the output to achieve higher entanglement. Our formalism makes a connection between nonclassicality of single-mode resources and NPT of output entangled states. 
The effect of PT in phase space is to change the sign of momentum,  $W_{\rho_{12}} ( q_{1}, p_{1}, q_{2}, p_{2} )\rightarrow W_{\rho_{12}} ( q_{1}, p_{1}, q_{2}, -p_{2} )$. If the resulting Wigner function is unphysical, the state $\rho_{12}$ is NPT.
We first show that all nonclassical states detected under our DMs can generate NPT entanglement via a BS setting.

We inject a single-mode state $\rho$ and its rotated version $\overline{\rho} = e^{i \frac{\pi}{2}\hat{n}} \rho e^{- i\frac{\pi}{2} \hat{n}}$ into a 50:50 beam splitter (BS), described as	
\begin{align}
		& W_{\rho} ( q_{1}, p_{1} ) W_{\overline{\rho}} ( q_{2}, p_{2} ) \nonumber \\
		\xrightarrow{\mathrm{BS}} & W_{\rho} ( \frac{q_{1} + q_{2}}{\sqrt{2}}, \frac{p_{1} + p_{2}}{\sqrt{2}} ) W_{\overline{\rho}} ( \frac{q_{1} - q_{2}}{\sqrt{2}},  \frac{p_{1} - p_{2}}{\sqrt{2}} ).
	\end{align}
Applying PT on mode 2 and injecting the state again into a 50:50 BS, we have
	\begin{align}
		& W_{\rho} ( \frac{q_{1} + q_{2}}{\sqrt{2}}, \frac{p_{1} - p_{2}}{\sqrt{2}} ) W_{\overline{\rho}} ( \frac{q_{1} - q_{2}}{\sqrt{2}},  \frac{p_{1} + p_{2}}{\sqrt{2}} ) \nonumber \\
		\xrightarrow{\mathrm{BS}} & W_{\rho} ( q_{1}, p_{2} ) W_{\overline{\rho}} ( q_{2}, p_{1} ) = W_{\rho} ( q_{1}, p_{2} ) W_{\rho} ( p_{1}, q_{2} ).
	\end{align}
Integrating over $q_{2}$ and $p_{2}$, the marginal Wigner function for mode 1 is given by $M_{\rho} ( q_{1} ) M_{\rho} ( p_{1} )$, which is identical to DM1 of the state $\rho$ in Eq.~\ref{eq:DM1}. The other DM2 in Eq.~\ref{eq:DM2} emerges when replacing the second input state $\overline{\rho}$ by a vacuum $\overline{\rho} = \ketbra{0}{0}$. Therefore, if the original state $\rho$ is nonclassical under our DMs, the output entangled state via the BS scheme must be NPT. 

In Ref. \cite{Asboth2005}, single-mode nonclassicality is characterized by entanglement potential via a BS setting, where a vacuum is used as an ancillary input to BS to generate entanglement. We may take negativity, instead of logarithmic negativity in \cite{Asboth2005}, as a measure of entanglement potential, i.e., 
	\begin{equation}
		\mathcal{P}_{\mathrm{ent}} [ \rho ] \equiv \frac{|| [ \hat{U}_{\mathrm{BS}} ( \rho_{1} \otimes \ketbra{0}{0}_{2} ) \hat{U}_{\mathrm{BS}}^{\dag} ]^{\mathrm{PT}} ||_{1} - 1}{2},
	\end{equation}
where $\hat{U}_{\mathrm{BS}}$ and $[\cdot]^{\mathrm{PT}}$ represent 50:50 beam-splitter operation and partial transpose on the mode 2, respectively. We then prove in \cite{supple} that our DM2  measure provides a lower bound for the entanglement potential as 
	\begin{equation} \label{eq:DMNO1}
		\mathcal{N}_{\mathrm{DM2}} [ \rho ] \leq \mathcal{P}_{\mathrm{ent}} [ \rho ].
	\end{equation} 
Thus the nonclassicality measured under our framework indicates the degree of entanglement achievable via BS setting.	

\section{Experiment}
We experimentally illustrate the power of our approach by detecting nonclassicality of several motional states of a trapped $^{171}$Yb$^{+}$ ion.	For the manipulation of motional state, the single phonon-mode $\hat{a}$ along $\rm X$-direction in 3-dimensional harmonic potential with trap frequencies $\left(\omega_{\rm X},\omega_{\rm Y}, \omega_{\rm Z}\right) = 2 \pi ~ (2.8,3.2,0.6)$ MHz 
is coupled to two internal levels of the $S_{1/2}$ ground state manifold, $|F=1,m_{F}=0\rangle\equiv|\uparrow\rangle$ and $|F=0,m_{F}=0\rangle\equiv|\downarrow\rangle$ with transition frequency $\omega_{\rm HF} = (2 \pi) ~ 12.642821$ GHz. 
We implement the anti-Jaynes-Cumming interaction $H_{\rm aJC} = \frac{\eta \Omega}{2} \hat{a}^{\dagger} \hat{\sigma}_{+} + {\rm h.c.}$ and the Jaynes-Cumming interaction $H_{\rm JC} = \frac{\eta \Omega}{2} \hat{a} \hat{\sigma}_{+} + {\rm h.c.}$ with $\sigma_{+}=|\uparrow\rangle\langle\downarrow|$. $H_{\rm aJC}$ is realized by two counter-propagating laser beams with beat frequency near $\omega_{\rm HF} + \omega_{\rm X}$ and $H_{\rm JC}$ with frequency near $\omega_{\rm HF} - \omega_{\rm X}$ \cite{Park2015}.  $\eta = \Delta k \sqrt{\hbar/2 M \omega_{\rm X} }$ is the Lamb-Dicke parameter, $\Omega$ the Rabi frequency of internal transition, $\Delta k$ the net wave-vector of the Raman laser beams and $M$ the ion mass.

For our test, we generate the Fock states $|n=1\rangle$ and $|n=2\rangle$, together with the ground state $|n=0\rangle$. First, we prepare the ground state by applying the standard Doppler cooling and the Raman sideband cooling. Then we produce the Fock states by a successive application of $\pi$-pulse of $H_{\rm aJC}$ transferring the state $|\downarrow, n\rangle$ to $|\uparrow, n+1\rangle$, and the $\pi$-pulse for internal state transition $|\uparrow, n+1\rangle$ to $|\downarrow, n+1\rangle$. We also generate a superposition state $\frac{1}{\sqrt{2}}(|0\rangle+|2\rangle)$ by applying the $\pi/2$ pulse of $H_{\rm aJC}$ and then the $\pi$ pulse of $H_{\rm JC}$.

{\bf Nonclassicality test}---
We measure a characteristic function $C_\rho(k_\theta)\equiv\langle e^{-2ik\hat x_\theta}\rangle$ with $\hat x_\theta=\hat x\cos\theta+\hat p \sin\theta$, by first making the evolution $\hat U=e^{-ik\hat x_\theta \hat\sigma_x}$ (simultaneously applying $H_{\rm aJC}$ and $H_{\rm JC}$ with proper phases) and then measuring internal state $\hat\sigma_z=|\uparrow\rangle\langle\uparrow|-|\downarrow\rangle\langle\downarrow|$ at times $t_i$ ($k=\eta\Omega t$) \cite{Roos, Roos'}. Using $\hat U^\dag \hat\sigma_z \hat U=\cos(2k\hat x_\theta) \hat\sigma_z+\sin(2k\hat x_\theta) \hat\sigma_y$, we obtain $\langle\cos(2k\hat x_\theta)\rangle$ and $\langle\sin(2k\hat x_\theta)\rangle$, with the internal state initially prepared in the eigenstates $|+\rangle_z$ and $|+\rangle_y$ of $\sigma_z$ and $\sigma_y$, respectively.
The Fourier transform of $C_\rho(k_\theta)$ gives the marginal distribution of $\hat x_\theta$ \cite{Roos, Roos'}. In contrast, we directly use it without the Fourier transform, for which our DMs work equally well as for the Wigner function.  We test $C^{\rm DM1}\equiv C_\rho(k_x)C_\rho(k_y)$ or $C^{\rm DM2}\equiv C_\rho(k_x)C_{|0\rangle\langle0|}(k_y)=C_\rho(k_x)e^{-\frac{1}{2}\lambda_{k_y}^2}$, with its density operator $\rho=\frac{1}{\pi}\int dk_xdk_y C(k_x,k_y){\hat D}^{\dag} (k_x,k_y)$ unphysical for a nonclassical state. 

\begin{figure}[!t]
	\centering
	\includegraphics[ scale = 0.62]{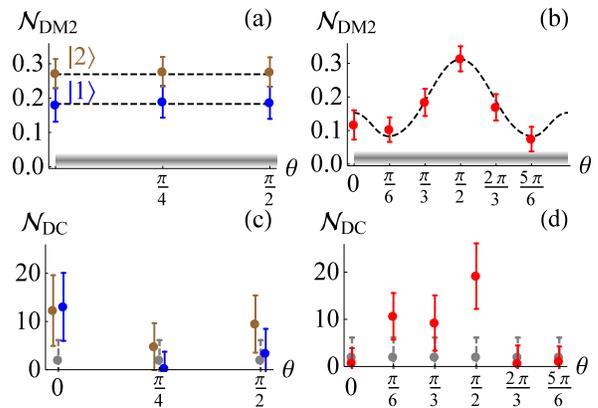}
		\caption{DM2 negativity against $\theta$ of the measured distribution $\langle e^{-2ik\hat x_\theta}\rangle$ for (a) Fock states $|1\rangle$ and $|2\rangle$ and (b) a superposition  $\frac{1}{\sqrt{2}}(|0\rangle+|2\rangle)$ . Dashed: theoretical value, bullets with error bars: experiment, Grey shade: noise level for classical states. (c)-(d): Negativity under deconvolution test for Fock states $|1\rangle$ (blue), $|2\rangle$ (brown), and $\frac{1}{2}(|0\rangle+|2\rangle)$ (red), together with $|0\rangle$ (gray) using a $5\times5$ moment-matrix. }
		\label{fig:NDM2EX}
	\end{figure}

To set a benchmark (noise level) for classical states, we prepared the motional ground state $|n=0\rangle$ and obtained its marginal distributions along six axes with 1000 repetitions for each time $t_i$. It yielded the negativity $\mathcal{N}_{\mathrm{DM2}}=0.019\pm0.02$ as represented by gray shading in Fig. \ref{fig:NDM2EX} \cite{supple}. On the other hand, the Fock states 
$|n=1\rangle$ and $|n=2\rangle$ clearly manifest nonclassicality for each marginal distribution taken at three different angles $\theta$ in Fig. 2 (a), at much higher negativity with error bars considering a finite data 1000. To further show that our method works regardless of measured axis, we also tested a superposition state $\frac{1}{\sqrt{2}}(|0\rangle+|2\rangle)$ not rotationally symmetric in phase space. As shown in Fig. 2 (b), its nonclassicality is well demonstrated for all measured angles individually while the degree of negativity varies with the measured axis.
	
Compared to our DM, one might look into nonclassicality directly via deconvolution, i.e. examining whether a marginal distribution $P(x)$ can be written as a sum of coherent-state distributions as $P(x)=\sqrt{\frac{2}{\pi}}\int d\tilde{x}\tilde{P}(\tilde{x})e^{-2(x-\tilde{x})^2}$, where $\tilde{P}(\tilde{x})$ must be positive-definite for classical states. $\tilde{P}(\tilde{x})$ is nothing but the marginal of Glauber-Sudarshan $P$-function, thus typically ill-behaved. One can test the positivity of $\tilde{P}(\tilde{x})$ alternatively using an $n\times n$ moment-matrix with elements $M_{ij}\equiv\langle \tilde{x}^{i+j}\rangle$ $(i,j=0,\dots,n-1) $\cite{Agarwal1993}. Figs. 2 (c) and (d) show the results under deconvolution using the same experimental data as we employed in Figs. 2 (a) and (b). To confirm nonclassicality, the degree of negativity must be large enough to beat that of the vacuum state including the statistical errors. Although those states produce negativity under deconvolution, their statistical errors are substantially overlapped with that of the vacuum state providing a much weaker evidence of nonclassicality than our DM. Full details are given in \cite{supple}.

Instead of employing an entire characteristic function, we can also test our criterion by examining a subset of data using the Kastler-Loupias-Miracle-Sole (KLM) condition \cite{KLM, KLM', KLM'', LegitimateW}. This simple test provides a clear evidence of nonclassicality against experimental imperfections e.g. coarse-graining and finite data acquisition in other experimental platforms as well. 
The KLM condition states that the characteristic function $C_{\rho} ( \xi ) \equiv \mathrm{tr} [ \rho \hat{D} ( \xi ) ]$ for a legitimate quantum state must yield a $n \times n$ positive matrix $\mathcal{M} > 0$ with matrix elements 
	\begin{equation} \label{eq:KLM}
		\mathcal{M}_{jk} = C ( \xi_{j} - \xi_{k} ) e^{\frac{1}{2} ( \xi_{j} \xi_{k}^{*} - \xi_{j}^{*} \xi_{k} )},
	\end{equation}
for an arbitrary set of complex variables $\{ \xi_{1}, \xi_{2}, ..., \xi_{n} \}$. 
In our case, we test the positivity of a matrix ($n=9$) constructed using $3\times3$ points of rectangular lattice of size $d$ for the characteristic function under DM2 \cite{supple}. As shown in Fig. 3 (a), the ground state $|0\rangle$ shows nonnegativity (thus the mixture of coherent states as well due to convexity of our method) for all values of $d$, whereas a nonclassical state $|\Psi\rangle=\frac{1}{\sqrt{2}}\left(|0\rangle+|2\rangle\right)$ manifests negativity in a certain range of $d$ confirming nonclassicality for each measured distribution at $\theta=\frac{\pi}{3},\frac{\pi}{2}$, and $\frac{2\pi}{3}$ (red solid curves) in Fig. 3 (b), (c) and (d), respectively. Furthermore, note that a mixture of the vacuum and the nonclassical state, $f|0\rangle\langle0|+(1-f)|\Psi\rangle\langle\Psi|$ possesses a positive-definite Wigner function for $f\ge0.66$, so even a full tomography may not directly show its nonclassicality via negativity. In contrast, our simple method manifests nonclassicality for $f=0.66$, as shown by blue dashed curves in Fig. 3 (b), (c) and (d). For Fock states, we consider the matrix test using 
$5\times5$ lattice-points, which confirms negativity at the mixing $f=0.5$ with vacuum giving a non-negative Wigner function for both states $|1\rangle$ and $|2\rangle$ in Fig. 3 (e) and (f). 

\begin{figure}[!t]
\centering
		\includegraphics[scale=0.62]{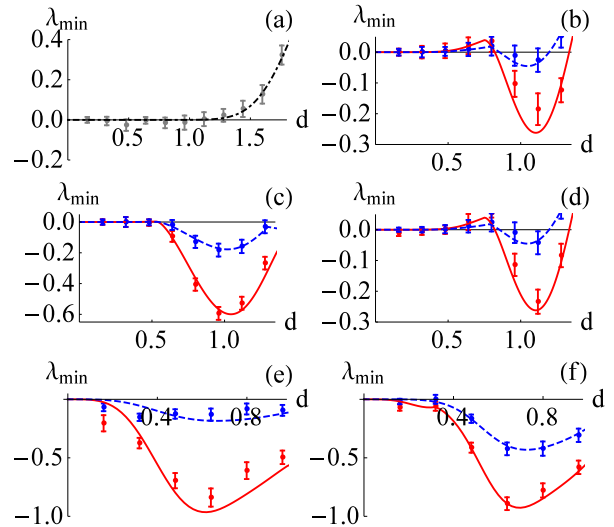}
	\caption{KLM test under DM2 using a matrix of $3\times3$ (a-d) and $5\times5$ (e,f) lattice points, respectively, with $\lambda_{\rm min}$ the lowest eigenvalue for each lattice size $d$ \cite{supple}. Negative $\lambda_{\rm min}$ manifests the nonclassicality of the considered state. Solid (pure-state $|\Psi\rangle$) and dashed (mixed-state $f|0\rangle\langle0|+(1-f)|\Psi\rangle\langle\Psi|$) curves represent theoretical predictions while bullets with error bars represent experimental data. (a) motional ground state $|0\rangle$, (b-d) $|\Psi\rangle=\frac{1}{\sqrt{2}}\left(|0\rangle+|2\rangle\right)$ for the measured angles at $\theta=\frac{\pi}{3},\frac{\pi}{2}$, and $\frac{2\pi}{3}$, respectively, (e) Fock state $|1\rangle$ and (f) Fock state $|2\rangle$. For mixed states (dashed curves), we use $f=0.66$ in (b-d) and $f=0.5$ in (e,f), respectively. }
		\label{fig:KLM3BY3}
\end{figure}

{\bf Genuine non-Gaussianity}---We further extend our approach combined with phase-randomization to derive a criterion on genuine non-Gaussianity. Notably there exist quantum tasks that cannot be achieved by Gaussian resources, e.g. universal quantum computation \cite{Lloyd}, CV nonlocality test \cite{NhaCV, NhaCV'}, entanglement distillation \cite{Eisert, Fiurasek, Giedke} and error correction \cite{Niset}. It is a topic of growing interest to detect genuine non-Gaussianity that cannot be addressed by a mixture of Gaussian states. Previous approaches all address particle nature like the photon-number distribution \cite{Filip2014, Jezek, Straka2014} and the number parity in phase space \cite{Park2015, Genoni2013,Hughes2014} for this purpose. Here we propose a method to examine a genuinely CV characteristics of marginal distributions. 
Our criterion can be particularly useful to test a class of non-Gaussian states diagonal in the Fock basis, $\rho=\sum p_n|n\rangle\langle n|$, thus rotationally symmetric in phase space. 
For this class, one may detect nonclassicality using photon-number moments \cite{Vogel'}, which can be experimentally addressed efficiently by phase-averaged quadrature measurements \cite{Munroe1995, Banaszek1997}.  Lvovsky and Shapiro experimentally demonstrated the nonclassicality of a noisy single-photon state $f|0\rangle\langle0|+(1-f)|1\rangle\langle1|$ for an arbitrary $f$ \cite{Shapiro} using the Vogel criterion \cite{Vogel''}. In contrast, we look into the genuine non-Gaussianity of non-Gaussian states as follows.

For a Gaussian state $\rho_{\rm G}$, the phase-randomization gives $\sigma\equiv\frac{1}{N} \sum_{k=0}^{N-1} e^{-i\theta_k{\hat n}}\rho_{\rm G}e^{i\theta_k{\hat n}}$ with $\theta_k\equiv\frac{k}{N}\pi$. As the number $N$ of phase rotation grows, the DM negativity of Gaussian states decreases. With $N\rightarrow\infty$ (full phase randomization), we obtain the Gaussian bound $\mathcal{B}_{\rm G} \approx 0.0887$ \cite{supple}. Thus, if a state manifests a larger DM negativity as $\mathcal {N}>\mathcal{B}_{\rm G}$, it confirms genuine non-Gaussianity. We plot the Gaussian bounds for finite rotations $N=6$ and $N=12$ with $\mathcal{B}_{\rm G} \approx 0.0887$ against energy $n$ in Fig. 4. Our data for the state $|2\rangle$, which shows negativity insensitive to measured angles in Fig. 2, indicates genuine non-Gaussianity for the mixed states $f|0\rangle\langle0|+(1-f)|2\rangle\langle2|$ with $f=1-\frac{n}{2}$. For example, the $N=12$ case (brown dot-dashed) as well as the full phase randomization (black dashed horizontal) confirms quantum non-Gaussianity at $f=\frac{1}{2}$ corresponding to a positive Wigner-function.

\begin{figure}[!t]
	\centering
	\includegraphics[scale=0.45]{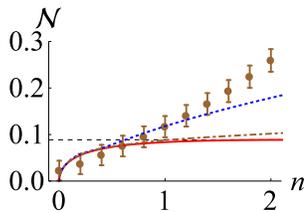}
	\caption{DM negativity of $f|0\rangle\langle0|+(1-f)|2\rangle\langle2|$ (bullet with error bar) from experimental data. Maximum Gausssian negativity under $N=6$ (blue dotted), $N=12$ (brown dot-dashed) and $N\rightarrow\infty$ (black dashed horizontal) phase rotations is given against energy $n$, the negativity above which confirms genuine non-Gaussianity.  }
	\label{fig:QNG}
\end{figure}


\section{Conclusion and Remarks}
Measuring marginal distributions along different axes in phase space forms a basis of quantum-state tomography with a wide range of applications. A marginal distribution is readily obtained in many different experimental platforms, e.g. by an efficient homodyne detection in quantum optical systems \cite{Lvovsky2009, Lvovsky2001, Ourjoumtsev2006, Huisman2009, Cooper2013} and by other quadrature measurements in trapped ion \cite{Roos, Roos', ion}, atomic ensembles \cite{as}, optomechanics \cite{opto, opto'}, and circuit QED systems \cite{CQED, CQED'}. We here demonstrated that only a single marginal distribution can manifest nonclassicality by employing our demarginalization maps (DMs). Our DM methods are powerful to detect a wide range of nonclassical states, particularly non-Gaussian states.  They provide a practical merit with less experimental efforts and make a stronger test of nonclassicality by analyzing data without numerical manipulation unlike state tomography.

Remarkably, nonclassicality can be demonstrated regardless of measured quadrature axis for all FDS states, which was also experimentally confirmed using a trapped ion system. We clearly showed that the proposed method provides a reliable nonclassicality test directly using a finite number of data, which can be further extended to other CV systems. In addition to the KLM test used here, we can manifest nonclassicality by looking into single marginal distributions under other forms, e.g. functional \cite{Nha2012} and entropic \cite{EUR, EUR'} inequalities.  
We also extended our approach to introduce a criterion on genuine non-Gaussianity employing marginal distributions combined with phase-randomization process. Our nonclassicality and non-Gaussianity tests were experimentally shown to successfully detect non-Gaussian states even with positive-definite Wigner functions whose nonclassicality is thus not immediately evident by the tomographic construction of Wigner function. As a note, for those nonclassical states with positive Wigner functions, one may use generalized quasi probability distributions like a filtered P-function  \cite{Agarwal1970-1,Agarwal1970-2,Agarwal1970-3}. For example, the experiment in \cite{Kiesel2011} introduced a nonclassicality filter to construct a generalized P-function that yields a regularized distribution with negativity as a signature of nonclassicality for the case of photon-added thermal states. On the other hand, our DM method does not require a tomographic construction and provides a faithful test reliable against experimental imperfections like finite data and coarse graining.

Moreover, we established the connection between single-mode nonclassicality and NPT entanglement via BS setting---a prototypical model of producing CV entanglement. The negativity under our DM framework provides a quantitative measure of useful resource by identifying the minimum level of entanglement achievable in Eq. 10 \cite{NN}. Nonclassicality and non-Gaussianity are important resources making a lot of quantum tasks possible far beyond classical counterparts. We thus hope our proposed method could provide a valuable experimental tool and a novel fundamental insight for future studies of CV quantum physics by critically addressing them.

{\it Acknowledgements}---
M.S.Z. and H.N. were supported by National Priorities Research Program Grant 8-751-1-157 from the Qatar National Research Fund and K.K. by the National Key Research and Development Program of China under Grants 2016YFA0301900 and 2016YFA0301901 and the National Natural Science Foundation of China under Grants 11374178 and 11574002.

\bibliographystyle{apsrev}

\vspace{1cm}
\hspace{2cm}{\bf Supplemental Information}
\section*{S1. Experiment: Analysis of fictitious characteristic function}
	
We here explain how our experimental data are analyzed to yield a quantitative measure of DM negativity for each state in Fig. 2 and the lowest eigenvalue of KLM test in Fig. 3 of main text.

\begin{itemize}
\item{\bf DM2 negativity}
\end{itemize}

\begin{figure}[!t]
		\includegraphics[scale=0.42]{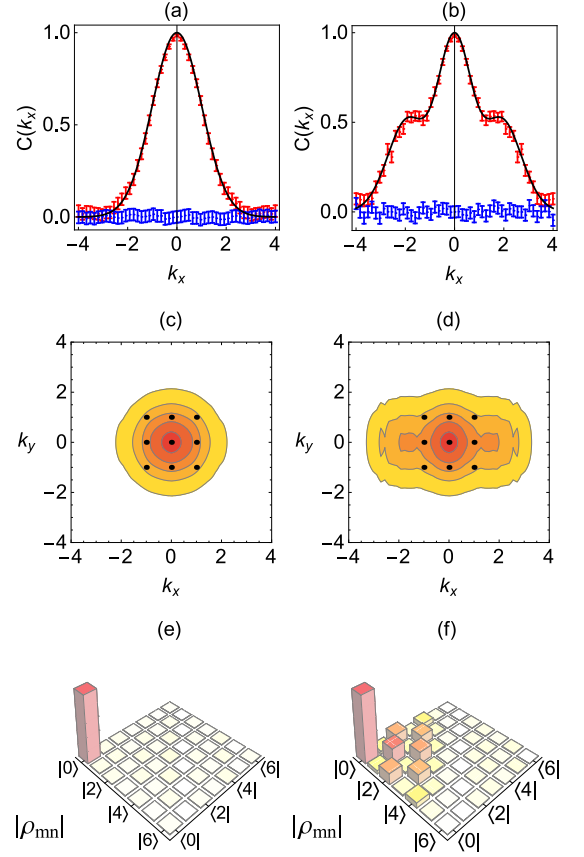}
		\caption{ A marginal distribution $C(k_x)$ of characteristic function for (a) ground state $|0\rangle$ and (b) superposition state $\frac{1}{\sqrt{2}}(|0\rangle+|2\rangle)$. Black solid curves represent theoretical predictions while red and blue dots represent the real and the imaginary parts of experimental data with error bars. (c), (d) Contour plots of fictitious characteristic function $C(k_x,k_y)=C(k_x)e^{-\frac{1}{2}k_y^2}$. (e), (f) matrix elements $\rho_{mn}^{\rm DM2}$ of the corresponding density operators. }
		\label{fig:CH1}
	\end{figure}

First, to determine the DM negativity (noise level) of classical states, we prepared the motional ground state $|0\rangle$ and observed the marginal distributions $C(k_{x_\theta})\equiv\langle e^{-2ik\hat x_\theta}\rangle$, where $\hat x_\theta=\hat x\cos\theta+\hat p \sin\theta$, over six different angles $\theta$ with 1000 data at each time $t_i$ with $k=\eta\Omega t_i$. The resulting distribution including all data with error bars at each $k_x$ is drawn in Fig.~\ref{fig:CH1} (a). A fictitious characteristic function is then constructed under DM2 as $C(k_x,k_y)=C(k_x)e^{-\frac{1}{2}k_y^2}$, of which contour plot is given in Fig.~\ref{fig:CH1} (c). We calculate the negativity of the corresponding density operator $\rho=\frac{1}{\pi}\int dk_xdk_y C(k_x,k_y){\hat D}^{\dag} (k_x,k_y)$ in the number-state basis. In Fig.~\ref{fig:CH1} (e), we show the matrix elements $\rho_{mn}^{\rm DM2}$ of the density operator. $\frac{|| \rho^{\mathrm{DM2}}||_{1} - 1}{2}$ gives the result $\mathcal{N}_{\mathrm{DM2}}=0.019\pm0.02$ in the main text.

On the other hand, for the case of a nonclassical state, e.g. $\frac{1}{\sqrt{2}}(|0\rangle+|2\rangle)$, the same procedures are taken for each measured axis, of which result, e.g. at $\theta=\frac{\pi}{2}$, is displayed in Fig.~\ref{fig:CH1} (b,d,f). 

\begin{itemize}
\item{\bf KLM test}
\end{itemize}
For the purpose of KLM test mentioned in the main text, we choose a matrix using $n\times n$ lattice points in the space of characteristic function. For example, we illustrate the case of $3\times 3$ square lattice (black dots) in Figs. ~\ref{fig:CH1} (c) and (d). We may look at matrices by changing the lattice size $d$ according to data availability, all of which must give nonnegative eigenvalues for classical states. In other words, if there exists a negative $\lambda_{\rm min}$ for any $d$, it confirms nonclassicality.

\section*{S2. Comparison of DM and deconvolution}
Given a marginal distribution, one can also look into nonclassicality via a deconvolution method instead of our DM. That is, one examines whether the obtained probability distribution $P(x)$ can be written as a convex sum of coherent-state distributions as $P(x)=\sqrt{\frac{2}{\pi}}\int d\tilde{x}\tilde{P}(\tilde{x})e^{-2(x-\tilde{x})^2}$, where $\tilde{P}(\tilde{x})$ must be positive-definite for classical states. In other words, if one finds negativity of $\tilde{P}(\tilde{x})$ under a Gaussian deconvolution of $P(x)$, it is a signature of nonclassicality. 
For their corresponding characteristic functions $C(k)=\int dx P(x)e^{-2ikx}$ and ${\tilde C}(k)=\int d\tilde{x} \tilde{P}(\tilde{x})e^{-2ik\tilde{x}}$, respectively, the deconvolution gives the relation ${\tilde C}(k)=C(k)e^{\frac{1}{2}k^2}$.

First, note that $\tilde{P}(\tilde{x})$ is nothing but the marginal distribution of Glauber-Sudarshan $P$-function, thus it is usually ill-behaved, e.g. delta-function for a coherent state and singular for Fock states. However, one can test the positivity of $\tilde{P}(\tilde{x})$ alternatively based on a moment test. For a classical state, a $n\times n$ matrix whose elements are given by $M_{ij}\equiv\langle \tilde{x}^{i+j}\rangle$ $(i,j=0,\dots,n-1)$ must be positive-definite at all levels of $n$. The moments $\langle \tilde{x}^{m}\rangle$ can be obtained via the moments $\langle x^{m}\rangle$ of the regular probability distribution $P(x)$ by way of 
$\langle \tilde{x}^{m}\rangle=2^{-3m/2}\langle H_m(\sqrt{2}x)\rangle$ with $H_m(x)$ the Hermite polynomial of order $m$. To derive it, let us use a method of operator ordering as follows. Due to the Baker--Campbell--Hausdorff formula, we have $e^{\lambda (\hat{a}+\hat{a}^{\dag})} = e^{\lambda^{2}/2} e^{\lambda \hat{a}^{\dag}} e^{\lambda \hat{a}}$, i.e. $:e^{\lambda ( \hat{a}+\hat{a}^{\dag} )}: = e^{\lambda \hat{a}^{\dag}} e^{\lambda \hat{a}} = e^{-\lambda^{2}/2} e^{\lambda (\hat{a}+\hat{a}^{\dag})}$.
We then obtain
	\begin{align}
		&:(\hat{a}+\hat{a}^{\dag})^{m}:=\lim_{\lambda \rightarrow 0} \frac{\partial^{m}}{\partial \lambda^{m}} :e^{\lambda ( \hat{a}+\hat{a}^{\dag} )} :\nonumber\\ & = \lim_{\lambda \rightarrow 0} \frac{\partial^{n}}{\partial \lambda^{m}} e^{-\lambda^{2}/2} e^{\lambda (\hat{a}+\hat{a}^{\dag})} \nonumber \\
		& = \lim_{\lambda \rightarrow 0} \sum_{k=0}^{m} \binom{m}{k} \frac{\partial^{k}}{\partial \lambda^{k}} e^{-\lambda^{2}/2} \frac{\partial^{m-k}}{\partial \lambda^{m-k}} e^{\lambda (\hat{a}+\hat{a}^{\dag})} \nonumber \\
		& = \lim_{\lambda \rightarrow 0} \sum_{k=0}^{m} \binom{m}{k} 2^{-k/2} e^{-\lambda^{2}/2} H_{k} ( - \lambda / \sqrt{2} ) ( \hat{a} + \hat{a}^{\dag} )^{m-k} e^{\lambda (\hat{a}+\hat{a}^{\dag})} \nonumber \\
		& = 2^{-m/2} \sum_{k=0}^{m} \binom{m}{k} H_{k} ( 0 ) \{ \sqrt{2} ( \hat{a} + \hat{a}^{\dag} ) \}^{m-k} \nonumber \\
		& = 2^{-m/2} H_{m} \bigg( \frac{\hat{a} + \hat{a}^{\dag}}{\sqrt{2}} \bigg),
	\end{align}
which yields the desired relation $: \hat{q}^{m} : = 2^{-3m/2} H_{m} ( \sqrt{2} \hat{q} )$ with $\hat{q} = \frac{\hat{a}+\hat{a}^{\dag}}{2}$. Note that we have used $H_{m} ( x ) = (-1)^{m} e^{x^{2}} \frac{d^{m}}{dx^{m}} e^{-x^{2}}$, $H_{m} (-x) = (-1)^{m} H_{m} (x)$ and $H_{m}(x+y) = \sum_{k=0}^{m} \binom{m}{k} H_{k}(x)(2y)^{m-k}$.

For a fair comparison with our DM, we here show the results of moment-matrix test under deconvolution for the states $|1\rangle$, $|2\rangle$ and  $\frac{1}{\sqrt{2}}(|0\rangle+|2\rangle)$ together with the vacuum $|0\rangle$ using the same experimental data as we employed in Fig. 2 of the main text. To confirm nonclassicality, the degree of negativity must be large enough to beat that of the vacuum state including the statistical errors. As shown in Fig. 6, those nonclassical states produce negativity under deconvolution, however their statistical errors are substantially overlapped with that of the vacuum state making the interpretation weak (with the only exception at $\theta=\frac{\pi}{2}$ for the state $\frac{1}{\sqrt{2}}(|0\rangle+|2\rangle)$). This is clearly contrasted with the results under our DM in Fig. 2 of the main text.

	\begin{figure}[!t]
	\includegraphics[scale=0.42]{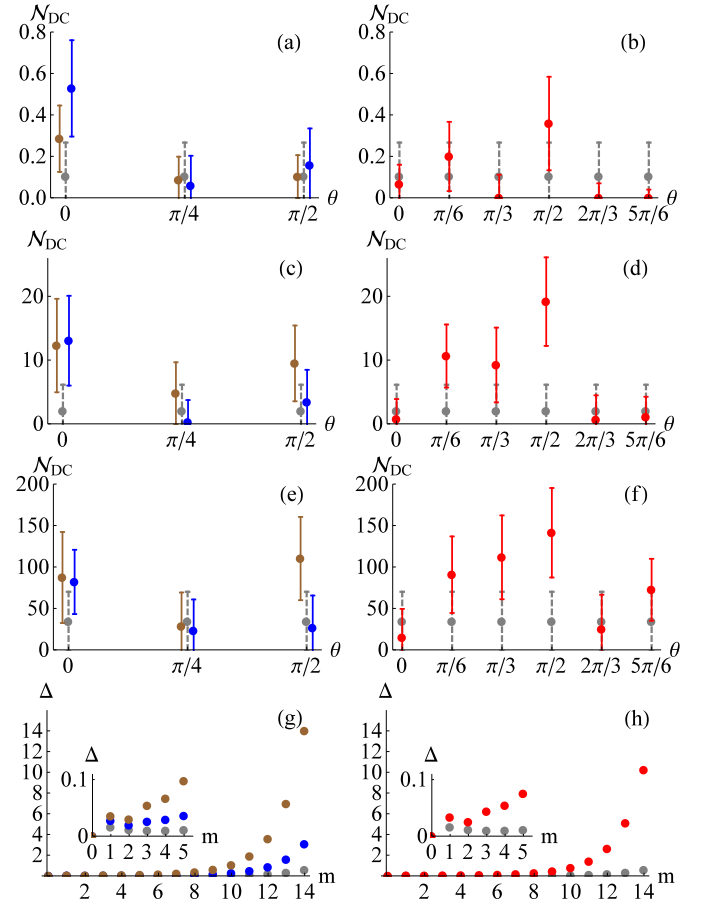}
		\caption{ (a)-(f) Negativity under deconvolution method with error bars for the same experimental data used in Fig. 2 of the main text. Left column: Fock states $|1\rangle$ (blue), $|2\rangle$ (brown), Right column: superposition state $\frac{1}{\sqrt{2}}(|0\rangle+|2\rangle)$ (red), together with $|0\rangle$ (gray) against measurement angle $\theta$. The size of the moment-matrix considered increases as $3\times3$ (a,b), $5\times5$ (c,d) and $7\times7$ (e,f), respectively. (g,h) Behavior of statistical error $\Delta_m$ of the moment $\langle \tilde{x}^{m}\rangle$ with a finite data $N=1000$ for (g) the Fock states $|1\rangle$ (blue), $|2\rangle$ (brown) and (h) the superposition $\frac{1}{\sqrt{2}}(|0\rangle+|2\rangle)$ (red) together with $|0\rangle$ (gray). Inset: magnified view for a small $m$.
 }
		\label{fig:DC}
	\end{figure}

One may try to optimize the matrix test by increasing the size $n$ of the matrix. This is theoretically valid, since the negativity not appearing in a low-dimension matrix can be found in a higher-dimension matrix as the latter encompasses the former. However, under practical situations with a finite number $N$ of data, the statistical error of the moment $\langle \tilde{x}^{m}\rangle$ given by $\Delta_m=\frac{2^{-3m/2}}{\sqrt N}\sqrt{\langle H_{m}^2 ( \sqrt{2} x )\rangle-\langle H_{m} ( \sqrt{2} x )\rangle^2}$ exponentially increase with $m$. For instance, $\Delta_m=\sqrt{m!4^{-m}/N}$ for the vacuum state whereas it grows at a higher rate for typical nonclassical states as shown in Fig. 6 (g) and (h). It is thus harder with a larger $m$ to beat the reference level (vacuum) of nonclassicality with significant errors, and the matrix test becomes optimized at a moderate level of matrix dimension. As shown in Fig. 6, the test does not necessarily improve by increasing the matrix size. If one obtains much more data $N$, the effect of statistical error may be reduced, however, the fact that our DM method works well already with a low number of data is an evidence of superioity in manifesting nonclassicality. Note that our DM method employs well-behaved square-integrable functions by its construction unlike the deconvolution method. Furthermore, unlike the deconvolution method aiming only at detecting nonclassicality, our DM formalism also constitute a useful framework to connect nonclassicality and quantum entanglement in CV setting and to address a genuine non-Gaussianity of CV systems that has been of growing interest.

\section*{S3. Testing Gaussian states under DMs}
As stated in the main text, a Gaussian nonclassical state, i.e. squeezed state, can be detected under our DMs if the measured marginal distribution is along a squeezed axis. For a given Gaussian state, we first determine a success range of measurement angles to manifest its nonclassicality. 
We later show that the nonclassicality can be demonstrated {\it regardless of measurement axis} by converting the Gaussian state to a non-Gaussian state under a finite number of phase rotations, which does not create nonclassicality. 

\begin{figure}[!t]
		\includegraphics[scale=0.7]{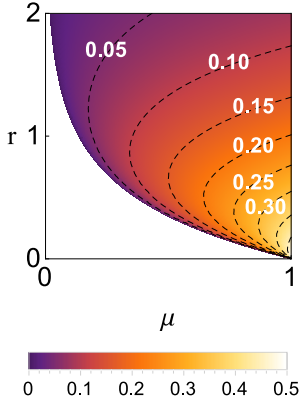}
		\caption{Contour for the probability of detecting nonclassicality for a Gaussian state with an unknown squeezing direction, i.e., $\eta_{\mathrm{sq}}$ in Eq.~\eqref{eq:ESQ}, against purity $\mu$ and squeezing $r$.}
		\label{fig:ESQ}
	\end{figure}

A single-mode Gaussian state can generally be represented as a displaced squeezed thermal state,
	\begin{equation} \label{eq:DSV}
		\sigma = \hat{D} ( \alpha ) \hat{S} ( r, \phi ) \sigma_{\mathrm{th}} ( \bar{n} ) \hat{S}^{\dag} ( r, \phi ) \hat{D}^{\dag} ( \alpha ),
	\end{equation}
where $\hat{S} ( r, \phi ) = \exp [- \frac{r}{2} ( e^{2i\phi} \hat{a}^{\dag 2} - e^{-2i\phi} \hat{a}^{2} ) ]$ is a squeezing operator ($r$: squeezing strength, $\phi$: squeezing direction), and $\sigma_{\mathrm{th}} ( \bar{n} ) = \sum_{n = 0}^{\infty} \frac{\bar{n}^{n}}{(\bar{n}+1)^{n+1}} \ketbra{n}{n}$ is a thermal state with mean photon number $\bar{n}$. 
The displacement does not affect the physicality issue under our DM methods, so we may set $\alpha = 0$ without loss of generality. Then a Wigner function with $\phi = 0$ (squeezed along $q$-axis) is given by
	\begin{equation}
		W_{\sigma} ( q, p ) = \frac{2 \mu}{\pi} \exp [ - 2 \mu ( e^{2r} q^{2} + e^{-2r} p^{2} ) ],
	\end{equation}
with $\mu = \frac{1}{1+2\bar{n}}$ the purity of the state. Its marginal distribution along the direction $\theta$ rotated from position $q$ is obtained as
	\begin{equation} \label{eq:GMD}
		M_{\sigma} ( x ) = \frac{1}{\sqrt{2 \pi V_{\theta}}} \exp \bigg( - \frac{x^{2}}{2V_{\theta}} \bigg),
	\end{equation}
where $V_{\theta} = \frac{1}{4 \mu} ( \cosh 2r - \cos 2\theta \sinh 2r )$ is the variance. We identify the squeezing range $V_{\theta} < \frac{1}{4}$ to be $| \theta| < \frac{1}{2} \arccos \frac{\cosh 2r - \mu}{\sinh 2r}\equiv | \theta |_{\mathrm{sq}} $. We thus obtain a success probability $\eta_{\mathrm{sq}}$ that a randomly chosen angle $\theta\in[0,\frac{\pi}{2}]$ for a marginal distribution can detect nonclassicality as
	\begin{equation} \label{eq:ESQ}
		\eta_{\mathrm{sq}} = \frac{| \theta |_{\mathrm{sq}}}{\frac{\pi}{2}}=\frac{1}{\pi} \arccos \frac{\cosh 2r - \mu}{\sinh 2r}.
	\end{equation}
We plot $\eta_{\mathrm{sq}}$ as a function of the purtiy $\mu \equiv \mathrm{tr} \rho^{2}$ and squeezing strength $r$ in Fig.~\ref{fig:ESQ}. 
In general, $\eta_{\mathrm{sq}}$ monotonically increases with the purity $\mu$. On the other hand, it decreases with $r$ as the range of squeezed quadratures decreases with the degree of squeezing. For instance, in the extreme squeezing $r\rightarrow\infty$, the range of squeezing direction becomes $| \theta |_{\mathrm{sq}}\rightarrow0$.

\begin{itemize}
\item{ \bf Detecting Gaussian states under a classicality-preserving operation}
\end{itemize}

We now introduce a finite number of phase rotations, which does not create nonclassicality, and show that it enables us to detect the nonclassicality of a Gaussian state regardless of measurement axis. The phase-randomizing operation generally transforms a Gaussian state to a non-Gaussian state and it was employed, e.g., for the distillation of squeezing \cite{Franzen} and the reconstruction of a Wigner function of rotationally-symmetric state (Fock states) \cite{Grangier}. We here use our DM2 approach and give proof in two-steps: (i) DM2 detects every nonclassical marginal distribution. (ii) The marginal distribution obtained from a squeezed state under phase-randomization is always nonclassical.

{\bf (i) Connection between DM2 and the Hamburger moment problem}

For a given state $\rho$, DM2 gives the Wigner function as
	\begin{equation}
		W_{\mathcal{D}_{2} [ \rho ]} (q,p) = M_{\rho}(q) \sqrt{\frac{2}{\pi}} e^{-2p^{2}},
	\end{equation}
which is equivalent to the Glauber-P representation as	
	\begin{equation}
		\mathcal{D}_{2} [ \rho ] = \int_{- \infty}^{\infty} dx T_{\rho} (x) \ketbra{x}{x}.
	\end{equation}
Here $\ket{x}$ represents a coherent state with real amplitude $x$ and the quasi-probability density $T_{\rho} (x)$ connects to the marginal distribution by $M_{\rho} ( q ) = \int_{-\infty}^{\infty} T_{\rho} (x) \sqrt{\frac{2}{\pi}} e^{-2(q-x)^{2}}$. (Note that $\mathcal{D}_{2} [ \rho ]$ is classical if and only if $T_{\rho} (x)$ is non-negative. We thus see that if the marginal distribution $M_{\rho} ( q )$ cannot be expressed as a positive sum of normal distributions with $\sigma=\frac{1}{2}$ (vacuum fluctuation), the corresponding $\mathcal{D}_{2} [ \rho ]$ cannot represent a classical state.)

In addition, the density operator $\mathcal{D}_{2} [ \rho ]$ is unphysical if and only if there exists a pure state $\ket{\psi} = \sum_{n} c_{n} \ket{n}$ satisfying
	\begin{align}
		\bra{\psi} \mathcal{D}_{2} [ \rho ] \ket{\psi} & = \sum_{n=0}^{\infty} \sum_{m = 0}^{\infty} c_{n}^{*} c_{m} \int_{- \infty}^{\infty} dx T_{\rho} (x) \braket{n}{x} \braket{x}{m} \nonumber \\
		& = \sum_{n=0}^{\infty} \sum_{m = 0}^{\infty} \frac{c_{n}^{*}}{\sqrt{n!}} \frac{c_{m}}{\sqrt{m!}} \int_{- \infty}^{\infty} dx T_{\rho} (x) e^{-x^{2}} x^{n+m} \nonumber \\
		& < 0.
	\end{align}
In measure theory, it is known that for the moments $a_{n} = \int_{- \infty}^{\infty} x^{n} d\rho (x)$, the measure $\rho$ is positive if and only if
	\begin{equation}
		\sum_{n = 0}^{N} \sum_{m = 0}^{N} \beta_{n}^{*} \beta_{m} a_{n+m} \geq 0,
	\end{equation}
is satisfied for every non-negative integer $N$ and every set of complex variables $\{ \beta_{0}, ... , \beta_{N} \}$ (Hamburger moment problem) \cite{ReedSimon}.
Comparing Eqs. (19) and (20), we see that $\mathcal{D}_{2} [ \rho ]$ is unphysical if and only if $T_{\rho} (x)$ fails to be positive. It therefore proves that $\mathcal{D}_{2} [ \rho ]$ becomes unphysical for every nonclassical marginal distribution.

{\bf (ii) Gaussian states under a finite number of phase rotations}

For a general Gaussian state in Eq. (13), we apply $N$ (finite) phase rotations of angles $\{ 0, \frac{1}{N} \pi, ..., \frac{N-1}{N} \pi \}$ with $N \geq \lceil \frac{1}{\eta_{\mathrm{sq}}} \rceil$, where $\eta_{\mathrm{sq}}$ is given in Eq. (16).  An equal mixture of those rotations gives a marginal distribution 
	\begin{equation}
		M_{\rho} ( x ) = \frac{1}{N} \sum_{k=0}^{N-1} \frac{1}{\sqrt{2 \pi v_{k}}} e^{-\frac{x^{2}}{2v_k}},
	\end{equation}
with $v_{k} = \frac{1}{4 \mu} [ \cosh 2r - \cos 2 ( \theta + \frac{k}{N} \pi ) \sinh 2r ]$. Importantly, there always exists a set of $k$ satisfying $v_{k} < \frac{1}{4}$ (squeezing) regardless of $\theta$ (measured axis), whenever $N \geq \lceil \frac{1}{\eta_{\mathrm{sq}}} \rceil$ is satisfied. Below we show that 
the distribution in Eq. (21) cannot be represented as a mixture of Gaussian distributions all with vacuum noise. Together with the property (i), this proves that the resulting non-Gaussian state can be detected regardless of quadrature axis under DM2 method.

Suppose that $M_{\rho} ( x )$ in Eq. (21) be written as a mixture of coherent-state distributions, i.e. 
\begin{eqnarray}
		M_{\rho} ( x ) = \frac{1}{N} \sum_{k=0}^{N-1} \frac{1}{\sqrt{2 \pi v_{k}}} e^{-\frac{x^{2}}{2v_k}}=\sum_{k'} \frac{p_{k'}}{\sqrt{\pi/2}} e^{-2(x-x_{k'})^{2}},\nonumber\\
	\end{eqnarray} 
then one encounters a contradiction. Let us take a Fourier transform, $\int dx e^{-ixy} M_{\rho} ( x )$, and then multiply both sides by $e^{\frac{v}{2}y^2}$ with $v\equiv{\rm min}\{v_k\}<\frac{1}{4}$. It gives  
\begin{eqnarray}
		\frac{1}{N}+ \frac{1}{N}\sum_{k\ne k_{\rm min}} e^{-\frac{1}{2}(v_k-v)y^2}=\sum_{k'} p_{k'}e^{-ix_{k'}y} e^{-\frac{1}{2}(\frac{1}{4}-v)y^2},\nonumber\\
	\end{eqnarray}
where the sum in the LHS excludes the terms with $v_k=v={\rm min}\{v_k\}$. It is readily seen that  Eq. (23) cannot be satisfied at all $y$, e.g., LHS=$\frac{1}{N}$ and RHS=0 as $y\rightarrow\infty$.

\begin{itemize}
\item{\bf Gaussian bound under phase randomization}
\end{itemize}

Let us now consider how the degree of nonclassicality (DM negativity) is bounded for all Gaussian states by a full phase-randomization. When the measured quadrature distribution is Gaussian with variance $V$, 
 the Wigner function under DM2 is given by 
	\begin{equation}
		W_{\sigma}^{\mathrm{DM2}} ( x, y ) = \frac{1}{\pi \sqrt{V}} \exp \bigg( - \frac{x^{2}}{2V} - 2y^{2} \bigg),
	\end{equation}
and the corresponding density matrix by
	\begin{equation}
		\sigma^{\prime} = \sum_{n = 0}^{\infty} \lambda_{n} \hat{S} ( r, \frac{\pi}{2} )  \ketbra{n}{n} \hat{S}^{\dag} ( r, \frac{\pi}{2} ),
	\end{equation}
where $r= - \frac{1}{2} \log 2 \sqrt{V}$ and $\lambda_{n} = \frac{2}{2\sqrt{V}+1} ( \frac{2\sqrt{V} - 1}{2\sqrt{V} + 1} )^{n}$ represent the eigenvalues of $\sigma^{\prime}$. A direct calculation gives its DM negativity as
	\begin{equation}
		N_{\mathrm{DM2}} ( V ) = \max \bigg( \frac{1}{4 \sqrt{V}} - \frac{1}{2}, 0 \bigg). 
	\end{equation}
Using the convexity of the DM negativity (proved in the next section), we deduce that the DM negativity of a squeezed state $\rho_{\rm G}$ under phase randomization, i.e. $\sigma\equiv\frac{1}{2\pi} \int_{0}^{2\pi} d\theta e^{-i\theta{\hat n}}\rho_{\rm G}e^{i\theta{\hat n}}$ must be bounded as
	\begin{align} \label{eq:GBU1}
		\mathcal{N}_{\mathcal{D} [ \sigma ]} & \leq \frac{1}{2\pi} \int_{0}^{2\pi} d\theta N_{\mathrm{DM2}} ( V_{\theta} ) \nonumber \\
		& = \frac{2}{\pi} \int_{0}^{\theta_{c}} d\theta N_{\mathrm{DM2}} \bigg( \frac{\cosh 2r - \sinh 2r \cos 2\theta}{4} \bigg) \nonumber \\
		& = \frac{1}{\pi} \left[ e^{r} \mathcal{F} ( \theta_{c}, 1 - e^{4r} ) - \theta_{c} \right],
	\end{align}
where $\mathcal{F} ( \phi, m ) = \int_{\theta = 0}^{\phi} ( 1 - m \sin^{2} \theta )^{-1/2} d \theta$ represents the elliptic integral of the first kind, and $\theta_{c} = \frac{1}{2} \arccos ( \tanh r )$ sets the boundary for squeezing $V_{\theta_{c}} = \frac{1}{4}$. 
While the DM negativity for a Gaussian state approaches infinity with squeezing, $\lim_{V \rightarrow 0} N_{\mathrm{DM2}} ( V ) = \infty$, the phase randomization reduces it to $\mathcal{B}_{\rm G} \approx 0.0887$, thus
	\begin{equation} \label{eq:GB}
		\mathcal{N}_{\mathcal{D} [ \sigma ]} \leq \mathcal{B}_{\rm G}.
	\end{equation}

\begin{figure}[!t]
		\includegraphics[scale=0.5]{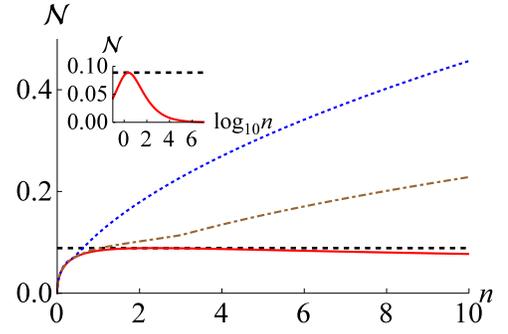}
		\caption{The DM negativity of a full phase-randomized Gaussian state is bounded by $\mathcal{B} \approx 0.0887$. The red solid line (inset as well) represents the upper bound of Eq. \eqref{eq:GBU1} while the blue dotted and the brown dot-dashed lines represent the cases of employing $N=6$ and $N=12$ phase rotations, respectively, against $n=\sinh^2r$.}
		\label{fig:PDQNG2}
	\end{figure}

\begin{figure}[!]
		\includegraphics[scale=0.5]{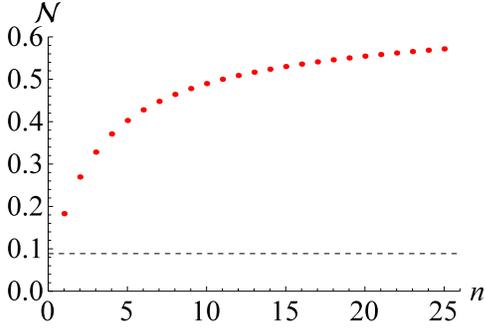}
		\caption{The DM negativity of each Fock state (red dot) is greater than the Gaussian bound $\mathcal{B} \approx 0.0887$ (black dashed).}
		\label{fig:PDQNG3}
	\end{figure}

We plot the upper bound of Eq.~\eqref{eq:GBU1} (red solid curve) in Fig.~\ref{fig:PDQNG2}, which takes maximum $\mathcal{B}_{\rm G} \approx 0.0887$. Due to the convexity of the DM negativity, we can further say that all Gaussian states and their mixtures also obey the same inequality. Therefore, if a state under a full phase-randomization violates the inequality as $\mathcal{N}_{\mathcal{D} [ \rho ]} > \mathcal{B}_{\rm G}$, it is a clear signature of genuine non-Gaussianity. 
For example, all Fock states achieve DM negativity above the Gaussian bound 
 as shown in Fig.~\ref{fig:PDQNG3}. 
Our method can detect genuine non-Gaussianity even for states having a non-negative Wigner function as shown in main text.

We can also numerically find the Gaussian bound for the case of finite $N$ phase-rotations, i.e. $\sigma'\equiv\frac{1}{N} \sum_{k=0}^{N-1} e^{-i\theta_k{\hat n}}\rho_{\rm G}e^{i\theta_k{\hat n}}$ with $\theta_k\equiv\frac{k}{N}\pi$. In Fig.~\ref{fig:PDQNG2}, the cases of $N=6$ and $N=12$ are plotted against energy $n$.

\section*{S4. Properties of DM negativity as a measure of nonclassicality}

Let us first define our DM negativity for each measured distribution as
	\begin{equation} \label{eq:DMN}
		\mathcal{N}_{\mathrm{DM}}^{\theta} ( \rho ) \equiv \frac{|| \rho_{\mathrm{DM}}^{\theta} ||_{1} - 1}{2},
	\end{equation}
where $||\cdot||_{1}$ represents a trace norm and $\rho_{\mathrm{DM}}^{\theta}$ the fictitious density matrix obtained by applying our DM method to the marginal distribution measured at angle $\theta$. It satisfies the following properties:
	\begin{enumerate}
		\item $\mathcal{N}_{\mathrm{DM}}^{\theta} ( \rho ) = 0$ for classical states.
		\item $\mathcal{N}_{\mathrm{DM}}^{\theta} ( \sum_{j} p_{j} \rho_{j} ) \leq \sum_{j} p_{j} \mathcal{N}_{\mathrm{DM}}^{\theta} ( \rho_{j} )$.
		\item $\mathcal{N}_{\mathrm{DM}}^{\theta} ( \hat{D} ( \alpha ) \rho \hat{D}^{\dag} ( \alpha ) ) = \mathcal{N}_{\mathrm{DM}}^{\theta} ( \rho )$.
		\end{enumerate}
		
\textbf{[P1]}
In the main text, we already showed that $\rho_{\mathrm{DM}}^{\theta}$ becomes a physical state when the initial state is classical, i.e. $|| \rho_{\mathrm{DM}}^{\theta} ||_{1} = 1$	 giving 
$\mathcal{N}_{\mathrm{DM}}^{\theta}=0 $.

\textbf{[P2]}
We can derive the convexity of DM negativity from the fact that the trace norm satisfies the homogeneity, i.e., $||cM||_{1} = |c| ||M||_{1}$, and the triangle inequality, i.e., $||M+N||_{1} \leq ||M||_{1} + ||N||_{1}$ \cite{Vidal2002, Plenio2002}.
	\begin{align*}
		\mathcal{N}_{\mathrm{DM}}^{\theta} ( \sum_{j} p_{j} \rho_{j} ) & = \frac{|| \sum_{j} p_{j} \rho_{j, \mathrm{DM}}^{\theta} ||_{1} - 1}{2} \nonumber \\
		& \leq \frac{\sum_{j} p_{j} || \rho_{j, \mathrm{DM}}^{\theta} ||_{1} - 1}{2} \nonumber \\
		& = \sum_{j} p_{j} \mathcal{N}_{\mathrm{DM}}^{\theta} ( \rho_{j} ).
	\end{align*}

\textbf{[P3]}
A displacement operation only shifts the center of Wigner function and marginal distributions.
	\begin{align}
		M_{\hat{D} ( \alpha ) \rho \hat{D}^{\dag} ( \alpha )} ( q_{\theta} ) & = \int dp_{\theta} W_{\hat{D} ( \alpha ) \rho \hat{D}^{\dag} ( \alpha )} ( q, p ) \nonumber \\
		& = \int dp_{\theta} W_{\rho} ( q - \mathrm{Re} [ \alpha ], p - \mathrm{Im} [ \alpha ] ) \nonumber \\
		& = M_{\rho} ( q_{\theta} - \mathrm{Re} [ \alpha e^{-i \theta} ] ).
	\end{align}
DM thereby yields a displaced version $\overline{\rho}_{DM}^{\theta} = \hat{D} ( \overline{\alpha} ) \rho_{DM}^{\theta} \hat{D}^{\dag} ( \overline{\alpha} )$ for the displaced state $\overline{\rho} = \hat{D} ( \alpha ) \rho \hat{D}^{\dag} ( \alpha )$ with $\overline{\alpha} = \mathrm{Re} [ \alpha e^{-i \theta} ]$, which therefore does not affect the physicality of the output density operator.
As the eigenspectrum of a Hermitian operator is invariant under a unitary operator, we conclude that $\mathcal{N}_{\mathrm{DM}}^{\theta} ( \hat{D} ( \alpha ) \rho \hat{D}^{\dag} ( \alpha ) ) = \mathcal{N}_{\mathrm{DM}}^{\theta} ( \rho )$.

While a phase rotation is also a classicality preserving operation, i.e., $e^{i \hat{n} \phi} \ket{\alpha} = \ket{\alpha e^{i\phi}}$, it can increase the value of $\mathcal{N}_{\mathrm{DM}}^{\theta} [ \rho ]$ as the degree of DM negativity depends on the measured angle $\theta$. Thus, to make the measure invariant under phase rotations as well, we introduce an optimized version of DM negativity as
	\begin{equation}
		\mathcal{N}_{\mathrm{DM}} ( \rho ) = \max_{\theta \in ( 0, \pi )} \mathcal{N}_{\mathrm{DM}}^{\theta} ( \rho ).
	\end{equation}
The DM negativity then satisfies 
	\begin{enumerate}
		\item $\mathcal{N}_{\mathrm{DM}} ( \rho ) = 0$ for classical states.

		\item $\mathcal{N}_{\mathrm{DM}} ( \sum_{j} p_{j} \rho_{j} ) \leq \sum_{j} p_{j} \mathcal{N}_{\mathrm{DM}} ( \rho_{j} )$.
		\item $\mathcal{N}_{\mathrm{DM}} ( \hat{U}_{c} \rho \hat{U}_{c}^{\dag} ) = \mathcal{N}_{\mathrm{DM}} ( \rho )$ where $\hat{U}_{c}$ is a classicality preserving unitary operation, i.e., displacement and phase rotation.
	\end{enumerate}

\textbf{[P1']}
It is a direct consequence of [P1].

\textbf{[P2']}
We also obtain the convexity of optimized DM negativity from [P2].
	\begin{align*}
		\mathcal{N}_{\mathrm{DM}} ( \sum_{j} p_{j} \rho_{j} ) & = \max_{\theta \in ( 0, \pi )} \mathcal{N}_{\mathrm{DM}}^{\theta} ( \sum_{j} p_{j} \rho_{j} ) \nonumber \\
		& \leq \max_{\theta \in ( 0, \pi )} \bigg( \sum_{j} p_{j} \mathcal{N}_{\mathrm{DM}}^{\theta} ( \rho_{j} ) \bigg) \nonumber \\
		& \leq \sum_{j} p_{j} \bigg( \max_{\theta \in ( 0, \pi )} \mathcal{N}_{\mathrm{DM}}^{\theta} ( \rho_{j} ) \bigg) \nonumber \\
		& = \sum_{j} p_{j} \mathcal{N}_{\mathrm{DM}} ( \rho_{j} ).
	\end{align*}

\textbf{[P3']}
The property [P3] implies that $\mathcal{N}_{\mathrm{DM}} ( \hat{D} ( \alpha ) \rho \hat{D}^{\dag} ( \alpha ) ) = \mathcal{N}_{\mathrm{DM}} ( \rho )$.  As a phase rotation only rotates the marginal distributions, i.e., $\mathcal{N}_{\mathrm{DM}}^{\theta} ( e^{-i \hat{n} \phi} \rho e^{i \hat{n} \phi} ) = \mathcal{N}_{\mathrm{DM}}^{\theta+\phi} ( \rho )$, the DM negativity optimized over the angles is not changed at all. Combining 2 and 3, we also deduce that \\

\indent 4. $\mathcal{N}_{\mathrm{DM}}$ does not increase under generic classicality preserving operations (mixture of unitary operations).\\

Let us here look into other nonclassicality measures in the literature considering the above properties.
Note that nonclassical depth \cite{Lee1991} does not satisfy the convexity. As an example, a mixture of vacuum and a squeezed state, i.e., $\sigma = (1-p) \ketbra{0}{0} + p \ketbra{\psi}{\psi}$, satisfies that $\tau ( \sigma ) = \tau ( \ketbra{\psi}{\psi} ) \neq 0$ for $0 < p \leq 1$ and $\tau ( \sigma ) = 0$ for $p = 0$.

Gehrke {\it et al.} addressed a set of conditions for a proper nonclassicality measure in \cite{Gehrke2012}. They have only addressed two conditions: (1) the measure is zero only for classical state. (2) the measure is nonincreasing for any classical Kraus operator. They also proposed a degree of nonclassicality which quantifies the number of superposition between coherent states required for representing the given state. As an example, the degree of nonclassicality for a pure state in the form $\sum_{j=0}^{d} c_{j} \ket{\alpha_{j}}$ is $d$. Similar to the case of nonclassical depth, the degree of nonclassicalty does not satisfy the convexity. For example, a mixture of vacuum ($d=0$) and a pure state $\ket{\psi}$ with non-zero degree $k$, $\sigma = (1-p) \ketbra{0}{0} + p \ketbra{\psi}{\psi}$, satisfies that $d ( \sigma ) = k$ for $0 < p \leq 1$ and $d ( \sigma ) = 0$ for $p = 0$. \\

{\bf Connection to entanglement potential}---
With the entanglement potential defined in the main text as 
	\begin{equation}
		\mathcal{P}_{\mathrm{ent}} [ \rho ] \equiv \frac{|| [ \hat{U}_{\mathrm{BS}} ( \rho_{1} \otimes \ketbra{0}{0}_{2} ) \hat{U}_{\mathrm{BS}}^{\dag} ]^{\mathrm{PT}} ||_{1} - 1}{2},
	\end{equation}
we here prove the relation 
\begin{equation} \label{eq:DMNO1}
		\mathcal{N}_{\mathrm{DM2}} [ \rho ] \leq \mathcal{P}_{\mathrm{ent}} [ \rho ].
	\end{equation} 
To this aim, we slightly modify the procedures in the main text as follows. After the beam-splitting operation for entanglement generation, we perform local unitary operations $e^{-i \theta \hat{n}_{1}} \otimes e^{-i \theta \hat{n}_{2}}$, which has no effect on entanglement and simply rotates the output Wigner function in Eq. (7) of main text as
	\begin{align}
		& W_{\rho} ( q_{+} \cos \theta - p_{+} \sin \theta, q_{+} \sin \theta + p_{+} \cos \theta ) \nonumber \\
		\times & W_{\ketbra{0}{0}} ( q_{-} \cos \theta - p_{-} \sin \theta, q_{-} \sin \theta + p_{-} \cos \theta ) \nonumber \\
		= & W_{\rho} ( q_{+} \cos \theta - p_{+} \sin \theta, q_{+} \sin \theta + p_{+} \cos \theta ) \nonumber \\
		\times & W_{\ketbra{0}{0}} ( q_{-}, p_{-} ),
	\end{align}
where $q_{\pm} = \frac{1}{\sqrt{2}} ( q_{1} \pm q_{2} )$, $p_{\pm} = \frac{1}{\sqrt{2}} ( p_{1} \pm p_{2} )$,  together with the fact that the Wigner function for vacuum is invariant under rotation. PT and BS operations change it to
	\begin{align}
		\xrightarrow{\mathrm{PT}}& W_{\rho} ( q_{+} \cos \theta - p_{-} \sin \theta, q_{+} \sin \theta + p_{-} \cos \theta ) \nonumber \\
		\times & W_{\ketbra{0}{0}} ( q_{-}, p_{+} ) \nonumber \\
		\xrightarrow{\mathrm{BS}} & W_{\rho} ( q_{1} \cos \theta - p_{2} \sin \theta, q_{1} \sin \theta + p_{2} \cos \theta ) \nonumber \\
		\times & W_{\ketbra{0}{0}} ( q_{2}, p_{1} ).
	\end{align}
Integrating over $q_{2}$ and $p_{2}$, we obtain the marginal Wigner function for one mode as $M_{\rho} ( x_{1} ) M_{\ketbra{0}{0}} ( y_{1} )$, which represents the demarginalization of the single mode state $\rho$ in Eq. (2) along with the rotated axis $x_{1} = q_{1} \cos \theta + p_{1} \sin \theta$. 

As a unitary operation preserves the eigenspectrum of an input state, we have
	\begin{equation}
		|| \rho^{\prime} ||_{1} = || \hat{U}_{\mathrm{BS}} \rho^{\prime} \hat{U}_{\mathrm{BS}}^{\dag} ||_{1},
	\end{equation}
where $\rho^{\prime} = [ \hat{R} ( \theta ) \hat{U}_{\mathrm{BS}} ( \rho_{1} \otimes \ketbra{0}{0}_{2} ) \hat{U}_{\mathrm{BS}}^{\dag} \hat{R}^{\dag} ( \theta ) ]^{\mathrm{PT}}$ and $\hat{R} ( \theta ) = e^{-i \theta \hat{n}_{1}} \otimes e^{-i \theta \hat{n}_{2}}$. Using the fact that trace norm is nonincreasing under partial trace \cite{Vidal2002, Plenio2002}, e.g., $|| \rho_{AB} ||_{1} \geq || \rho_{A} ||_{1}$, we obtain
	\begin{equation}
		|| \rho^{\prime} ||_{1} \geq || \mathrm{Tr}_{2} [ \hat{U}_{\mathrm{BS}} \rho^{\prime} \hat{U}_{\mathrm{BS}}^{\dag} ] ||_{1},
	\end{equation}
which is satisfied for every rotation angle $\theta$. From the obtained inequality, we derive the relation in Eq.~\eqref{eq:DMNO1}.

Furthermore, we can also show that two measures are identical for a Gaussian state, $\mathcal{N}_{\mathrm{DM2}} [  \sigma ] = \mathcal{P}_{\mathrm{ent}} [ \sigma ]$ by direct calculations.

\section*{S5. All FDS can be detected via a single marginal distribution regardless of measurement axis.}
\begin{widetext}
\begin{itemize}
\item{\bf Derivation of Wigner function for FDS}
\end{itemize}

The Wigner function of the FDS $\rho = \sum_{j,k = 0}^{N} \rho_{jk} \ketbra{j}{k}$ is given by
	\begin{equation}
		W_{\rho} ( x, y ) = \sum_{j,k = 0}^{N} \rho_{jk} W_{\ketbra{j}{k}} ( x, y ),
	\end{equation}
where $W_{\ketbra{j}{k}} ( x, y )$ represents the operator $\ketbra{j}{k}$
	\begin{align} \label{eq:WF}
		W_{\ketbra{j}{k}} ( x, y ) = \frac{2}{\pi} e^{- 2 | \alpha_{\theta} |^{2}} \sqrt{\frac{k!}{j!}} (-1)^{k} ( 2 \alpha_{\theta}^{*} )^{j-k} L_{k}^{(j-k)} ( 4 | \alpha_{\theta} |^{2} )\hspace{0.1cm} (j \geq k),
	\end{align}
with $L_{n}^{(\alpha)} ( z )$ a generalized Laguerre polynomial of order $n$, $\alpha_{\theta} = ( x + iy ) e^{-i \theta}$ and $W_{\ketbra{j}{k}} ( x, y ) = W_{\ketbra{k}{j}}^* ( x, y )$ for $j < k$, as shown in Ref. \cite{UL}. With $\alpha_{\theta=0} = q+ip$, we have a simple relation $W_{\ketbra{j}{k}} ( x, y )=e^{i\theta(j-k)}W_{\ketbra{j}{k}} ( q, p )$, thus we are only concerned with $W_{\ketbra{j}{k}} ( q, p )$ letting $\theta=0$.

We first reexpress Eq.~\eqref{eq:WF} using $\frac{d^{k}}{dx^{k}} L_{n}^{( \alpha )} ( x ) = (-1)^{k} L_{n-k}^{( \alpha + k )} ( x )$,
	\begin{align} \label{eq:WH1}
		W_{\ketbra{j}{k}} ( q, p ) & = \frac{2}{\pi} e^{-2q^{2}-2p^{2}} \sqrt{\frac{k!}{j!}} (-1)^{k} (2q-2ip)^{j-k} L_{k}^{(j-k)} ( 4q^{2}+4p^{2} ) \nonumber \\
		& = \frac{2}{\pi} e^{-2q^{2}-2p^{2}} \sqrt{\frac{k!}{j!}} \frac{(-1)^{j}}{2^{j-k}} \bigg[ \frac{\partial^{j-k}}{\partial \alpha^{j-k}} L_{j} ( 4 \alpha \alpha^{*} ) \bigg]_{\alpha = q + ip}.
	\end{align}
Employing the formulas in Ref.~\cite{AS}, i.e. $L_{n}^{( \alpha + \beta + 1 )} ( x + y ) = \sum_{j = 0}^{n} L_{j}^{( \alpha )} ( x ) L_{n - j}^{( \beta )} ( y )$ and $H_{2n} ( x ) = (-4)^{n} n! L_{n}^{(-1/2)} ( x^{2} )$,
we obtain
	\begin{align} \label{eq:WH2}
		L_{j} ( 4 q^{2} + 4 p^{2} ) & = \sum_{\ell = 0}^{j} L_{\ell}^{(-1/2)} ( 4q^{2} ) L_{j - \ell}^{(-1/2)} ( 4p^{2} ) \nonumber \\
		& = \sum_{\ell = 0}^{j} \frac{H_{2 \ell} ( 2q ) H_{2 ( j - \ell )} ( 2p )}{(-4)^{j} \ell ! ( j - \ell )!} .
	\end{align}
Combining Eqs. \eqref{eq:WH1} and \eqref{eq:WH2} with $\frac{d}{dx} H_{n} ( x ) = 2n H_{n-1} ( x )$ and $\frac{d^{n+1}}{dx^{n+1}} H_{n} ( x )=0$, we obtain
	\begin{align} \label{eq:WH3}
		W_{\ketbra{j}{k}} ( q, p ) & = \frac{2}{\pi} e^{-2q^{2}-2p^{2}} \sqrt{\frac{k!}{j!}} \frac{(-1)^{j}}{2^{j-k}} \bigg[ \frac{\partial^{j-k}}{\partial \alpha^{j-k}} \sum_{\ell = 0}^{j} \frac{H_{2 \ell} ( \alpha^{*} + \alpha ) H_{2 ( j - \ell )} ( i \alpha^{*} - i \alpha )}{(-4)^{j} \ell ! ( j - \ell )!} \bigg]_{\alpha = q + ip} \nonumber \\
		& = \frac{2}{\pi} e^{-2q^{2}-2p^{2}} \sqrt{\frac{k!}{j!}} \frac{(-1)^{j}}{2^{j-k}} \sum_{\ell = 0}^{j} \bigg[ \sum_{m = s}^{t} \frac{1}{(-4)^{j} \ell ! ( j - \ell )!} \frac{(j-k)!}{m! (j-k-m)!} \frac{2^{m} (2\ell)!}{(2\ell - m)!} H_{2 \ell - m} ( \alpha^{*} + \alpha ) \nonumber \\
		& \times \frac{(-2i)^{j - k - m} \{2 (j-\ell) \}!}{\{ 2 (j-\ell) - (j-k-m) \}!} H_{2 (j-\ell) - (j-k-m)} ( i \alpha^{*} - i \alpha ) \bigg]_{\alpha = q + ip} \nonumber \\
		& = \frac{2}{\pi} e^{-2q^{2}-2p^{2}} \sqrt{\frac{k!}{j!}} \frac{1}{4^{j}} \frac{(j-k)!}{j!} \sum_{\ell = 0}^{j} \sum_{m = s}^{t} \binom{j}{\ell} \binom{2\ell}{m} \binom{2(j-\ell)}{j-k-m} \frac{H_{2 \ell - m} ( 2q ) H_{j+k+m-2\ell} ( 2p )}{i^{j-k-m}},
	\end{align}
with $s = \max [ 0, 2\ell - j-k ]$ and $t = \min[j-k,2\ell]$, where we used $\frac{d^n}{dx^n} [f( x )g(x)]=\sum_{s=0}^n \frac{n!}{(n-s)!s!}\frac{d^{n-s}f}{dx^{n-s}}\frac{d^sg}{dx^s}$. Finally, we can recast Eq.~\eqref{eq:WH3} to
	\begin{equation}\label{eq:W_jk}
		W_{\ketbra{j}{k}} ( q, p ) = \frac{2}{\pi} e^{-2q^{2}-2p^{2}} \sum_{n = 0}^{j+k} A_{\ketbra{j}{k}} ( n ) H_{n} (2q) H_{j+k-n} (2p), 
	\end{equation}
where
	\begin{align}\label{eq:AE}
		A_{\ketbra{j}{k}} ( n ) & = \sqrt{\frac{k!}{j!}} \frac{1}{4^{j}} \frac{(j-k)!}{j!} \sum_{\ell = 0}^{j} \sum_{m = s}^{t} \binom{j}{\ell} \binom{2\ell}{m} \binom{2(j-\ell)}{j-k-m} \frac{\delta_{n, 2\ell-m}}{i^{j-k-m}} \nonumber \\
		& =
		\begin{cases}
			\displaystyle \sqrt{\frac{k!}{j!}} \frac{1}{4^{j}} \frac{(j-k)!}{j!} \sum_{\ell = \lceil \frac{n}{2} \rceil}^{\lfloor \frac{j-k+n}{2} \rfloor} \binom{j}{\ell} \binom{2\ell}{2\ell-n} \binom{2(j-\ell)}{j-k+n-2\ell} \frac{(-1)^{\ell}}{i^{j-k+n}} & \mbox{for $j>k$,} \\
			\displaystyle \frac{1}{4^{j} (\frac{n}{2})! (j-\frac{n}{2})!} & \mbox{for $j=k$ and even $n$,} \\
			\displaystyle 0 & \mbox{for $j=k$ and odd $n$.}
		\end{cases}
	\end{align}
We also provide an alternative expression of $A_{\ketbra{j}{k}} ( n )$ for even $j+k-n$ as
	\begin{equation} \label{eq:AEA}
		A_{\ketbra{j}{k}} ( n ) = \frac{(-1)^{\frac{j+k-n}{2}} \sqrt{j! k!}}{2^{j+k} n! (j+k-n)!} \sum_{r = 0}^{n} (-1)^{k-r} \binom{n}{r} \binom{j+k-n}{k-r}.
	\end{equation}
This can be obtained by comparing two methods of deriving marginal distribution for $\ketbra{j}{k}$, that is, $M_{\ketbra{j}{k}} ( q ) = \int dp W_{\ketbra{j}{k}} ( q, p )$ and $M_{\ketbra{j}{k}} ( q ) = \braket{q}{j} \braket{k}{q}$, where $\ket{q}$ is an eigenstate of the position operator, $\hat{q} \ket{q} = q \ket{q}$. Using the formula in Ref. \cite{Wang2008}, $H_{j} ( x ) H_{k} ( x ) = \sum_{n = 0}^{j+k} a_{jk} (n) H_{n} ( \sqrt{2} x )$ with 
	\begin{equation}
		a_{jk} (n) =
			\begin{cases}
				\frac{(-1)^{\frac{j+k-n}{2}} j! k!}{2^{\frac{j+k}{2}} n! ( \frac{j+k-n}{2} )!} \sum_{r = 0}^{n} (-1)^{k-r} \binom{n}{r} \binom{j+k-n}{k-r} & \mbox{for even $j+k-n$,} \\
				0 & \mbox{for odd $j+k-n$,}
			\end{cases}
	\end{equation}
and the expression in Ref. \cite{BR}, $\braket{q}{n} = ( \frac{2}{\pi} )^{\frac{1}{4}} \sqrt{\frac{1}{2^{n} n!}} e^{-q^{2}} H_{n} ( \sqrt{2} q )$, we obtain
	\begin{align} \label{eq:c1}
		\braket{q}{j} \braket{k}{q} & = \sqrt{\frac{2}{\pi}} \frac{1}{2^{\frac{j+k}{2}} \sqrt{j! k!}} e^{-2q^{2}} H_{j} ( \sqrt{2} q ) H_{k} ( \sqrt{2} q ), \nonumber \\
		& = \sqrt{\frac{2}{\pi}} e^{-2q^{2}} \sum_{n = 0}^{j+k} \frac{a_{jk} ( n )}{2^\frac{j+k}{2} \sqrt{j! k!}} H_{n} ( 2q ).
	\end{align}
On the other hand, from Eq. \eqref{eq:W_jk}, we obtain
	\begin{align} \label{eq:c2}
		\int_{-\infty}^{\infty} dp W_{\ketbra{j}{k}} ( q, p ) & =
		\begin{cases}
		 \sqrt{\frac{2}{\pi}} e^{-2q^{2}} \sum_{m = 0}^{\frac{j+k}{2}} \frac{(j+k-2m)!}{(\frac{j+k}{2}-m)!} A_{\ketbra{j}{k}} (2m) H_{2m} (2q) & \mbox{for even $j+k$,} \\
				\sqrt{\frac{2}{\pi}} e^{-2q^{2}} \sum_{m = 0}^{\frac{j+k-1}{2}} \frac{(j+k-2m-1)!}{(\frac{j+k-1}{2}-m)!} A_{\ketbra{j}{k}} (2m+1) H_{2m+1} (2q) & \mbox{for odd $j+k$,}
			\end{cases}
		\nonumber \\
		& = \sqrt{\frac{2}{\pi}} e^{-2q^{2}} \sum_{m^{\prime} = 0}^{\left[\frac{j+k}{2}\right]} \frac{(2m^{\prime})!}{m^{\prime}!} A_{\ketbra{j}{k}} (j+k-2m^{\prime}) H_{j+k-2m^{\prime}} (2q),
	\end{align}
where we have used $\int_{-\infty}^{\infty} dx e^{-2x^{2}} H_{2m} ( 2x ) = \sqrt{\frac{\pi}{2}} \frac{(2m)!}{m!}$ and $\int_{-\infty}^{\infty} dx e^{-2x^{2}} H_{2m+1} ( 2x ) = 0$ for an integer $m\geq0$. Comparing Eqs. \eqref{eq:c1} and \eqref{eq:c2}, we obtain Eq. \eqref{eq:AEA}. 

\begin{itemize}
\item{\bf Unphysicality of all FDSs under DM1 and DM2 methods}
\end{itemize}

Starting with a FDS state $\rho = \sum_{j,k = 0}^{N} \rho_{jk} \ketbra{j}{k}$ with its highest excitation $N$, i.e. $\rho_{NN}\ne0$, the density matrix elements for $W_{\rho}^{\mathrm{DM1}} ( x, y )$ under DM1 are obtained by
	\begin{align}
		\rho_{n_{1}, n_{2}}^{\prime} & = \pi \int_{- \infty}^{\infty} dx \int_{- \infty}^{\infty} dy W_{\rho}^{\mathrm{DM1}} ( x, y ) W_{\ketbra{n_{2}}{n_{1}}} ( x, y ) \nonumber \\
		& = \pi \int_{- \infty}^{\infty} dx \int_{- \infty}^{\infty} dy \bigg( \frac{2}{\pi} \bigg)^{2} e^{-4x^{2}-4y^{2}} \sum_{j,k = 0}^{N} \rho_{jk} e^{i(j-k)\theta} \sum_{j^{\prime},k^{\prime} = 0}^{N}\rho_{j^{\prime}k^{\prime}} e^{i(j^{\prime}-k^{\prime})\theta} \nonumber \\
		& \times \sum_{m=0}^{\left[\frac{j+k}{2} \right]} \frac{(2m)!}{m!} A_{\ketbra{j}{k}} (j+k-2m) H_{j+k-2m} (2x) \sum_{m^{\prime}=0}^{\left[\frac{j^{\prime}+k^{\prime}}{2} \right]} \frac{(2m^{\prime})!}{m^{\prime}!} A_{\ketbra{j^{\prime}}{k^{\prime}}} (j^{\prime}+k^{\prime}-2m^{\prime}) H_{j^{\prime}+k^{\prime}-2m^{\prime}} (2y) \nonumber \\
		& \times \sum_{n=0}^{n_{1}+n_{2}} A_{\ketbra{n_{2}}{n_{1}}} ( n ) H_{n} (2x) H_{n_{1}+n_{2}-n} (2y).
	\end{align}
Using the orthogonality of Hermite polynomials, $\int dx e^{-4x^{2}} H_{a} (2x) H_{b} (2x) = \sqrt{\pi} 2^{b-1} b! \delta_{a,b}$, we have
	\begin{align}
		\rho_{n_{1}, n_{2}}^{\prime} & = \sum_{j,k = 0}^{N}\rho_{j k} e^{i(j-k)\theta} \sum_{j^{\prime},k^{\prime} = 0}^{N}\rho_{j^{\prime}k^{\prime}} e^{i(j^{\prime}-k^{\prime})\theta} \sum_{m=0}^{\left[\frac{j+k}{2} \right]} \sum_{m^{\prime}=0}^{\left[\frac{j^{\prime}+k^{\prime}}{2} \right]} \sum_{n=0}^{n_{1}+n_{2}} \delta_{n, j+k-2m} \delta_{n_{1}+n_{2}-n,j^{\prime}+k^{\prime}-2m^{\prime}} \nonumber \\
		& \times 2^{n_{1}+n_{2}} n! (n_{1}+n_{2}-n)! \frac{(2m)!}{m!} \frac{(2m^{\prime})!}{m^{\prime}!} A_{\ketbra{j}{k}} (n) A_{\ketbra{j^{\prime}}{k^{\prime}}} (n_{1}+n_{2}-n) A_{\ketbra{n_{2}}{n_{1}}} ( n ).
	\end{align}
Looking into Kronecker delta functions and $j+k+j^{\prime}+k^{\prime} \leq 4N$, we first have $\rho_{n_{1}, n_{2}}^{\prime} = 0$ for $n_{1} + n_{2} > 4N$ that naturally imposes $\rho^{\prime}_{n,n} = 0$ for $n > 2N$. We now derive compact expressions for $\rho_{4N,0}^{\prime}$ and $\rho_{0,4N}^{\prime}$, which correspond to $j=k=j'=k'=N$ and $m=m'=0$, as
	\begin{equation}
		\rho_{4N,0}^{\prime} = \rho_{0,4N}^{\prime} = \rho_{NN}^{2} 16^{N} \{(2N)!\}^{2} A_{\ketbra{N}{N}}(2N)^{2} A_{\ketbra{4N}{0}} (2N) = \frac{\sqrt{(4N)!}}{(-16)^N (N!)^{2}} \rho_{NN}^{2},
	\end{equation}
where we have used $A_{\ketbra{N}{N}} (2N) = \frac{1}{4^{N} N!}$ and $A_{\ketbra{4N}{0}} (2N) = \frac{\sqrt{(4N)!}}{(-16)^{n} \{(2N)!\}^{2}}$ from Eqs. \eqref{eq:AE} and \eqref{eq:AEA}. For the determinant of a sub-matrix, we find that
	\begin{equation}
		\rho_{0,0}^{\prime} \rho_{4N,4N}^{\prime} - \rho_{4N,0}^{\prime} \rho_{0,4N}^{\prime} = - \frac{(4N)!}{256^{N} (N!)^{4}} \rho_{NN}^{4} < 0,
	\end{equation}
which confirms the nonclassicality of the original FDS state $\rho = \sum_{j,k = 0}^{N} \rho_{jk} \ketbra{j}{k}$. 

Similarly, the density matrix elements for $W_{\rho}^{\mathrm{DM2}} (x,y)$ under DM2 are given by
	\begin{align}
		\rho_{n_{1}, n_{2}}^{\prime\prime} & = \pi \int_{- \infty}^{\infty} dx \int_{- \infty}^{\infty} dy W_{\rho}^{\mathrm{DM2}} ( x, y ) W_{\ketbra{n_{2}}{n_{1}}} ( x, y ) \nonumber \\
		& = \pi \int_{- \infty}^{\infty} dx \int_{- \infty}^{\infty} dy \bigg( \frac{2}{\pi} \bigg)^{2} e^{-4x^{2}-4y^{2}} \sum_{j,k = 0}^{N} \rho_{jk} e^{i(j-k)\theta} \nonumber \\
		& \times \sum_{m=0}^{[ \frac{j+k}{2} ]} \frac{(2m)!}{m!} A_{\ketbra{j}{k}} (j+k-2m) H_{j+k-2m} (2x) \sum_{n=0}^{n_{1}+n_{2}} A_{\ketbra{n_{2}}{n_{1}}} ( n ) H_{n} (2x) H_{n_{1}+n_{2}-n} (2y) \nonumber \\
		& = \sum_{j,k = 0}^{N}\rho_{jk} e^{i(j-k)\theta} \sum_{m=0}^{[ \frac{j+k}{2} ]} \sum_{n=0}^{n_{1}+n_{2}} \delta_{n, j+k-2m} \delta_{n_{1}+n_{2}-n,0} 2^{n_{1}+n_{2}} n! \frac{(2m)!}{m!} A_{\ketbra{j}{k}} (n) A_{\ketbra{n_{2}}{n_{1}}} ( n ) \nonumber \\
		& = \sum_{j,k = 0}^{N}\rho_{jk} e^{i(j-k)\theta} \sum_{m=0}^{[ \frac{j+k}{2} ]} \delta_{n_{1}+n_{2}, j+k-2m} 2^{n_{1}+n_{2}} (n_{1}+n_{2})! \frac{(2m)!}{m!} A_{\ketbra{j}{k}} (n_{1}+n_{2}) A_{\ketbra{n_{2}}{n_{1}}} (n_{1}+n_{2}),
	\end{align}
which manifests that $\rho_{n_{1}, n_{2}}^{\prime\prime} = 0$ for $n_{1}+n_{2}>2N$. In addition, we obtain compact expressions for $\rho_{2N,0}^{\prime\prime}$ and $\rho_{0,2N}^{\prime\prime}$, for which $j=k=N$ and $m=0$, as
	\begin{equation}
		\rho_{2N,0}^{\prime\prime} = \rho_{0,2N}^{\prime\prime} = \rho_{NN} 2^{2N} (2N)! A_{\ketbra{N}{N}} (2N) A_{\ketbra{2N}{0}} (2N) = \frac{\sqrt{(2N)!}}{4^{N} N!} \rho_{NN},
	\end{equation}
where $A_{\ketbra{2N}{0}} (2N) = \frac{1}{4^{N} \sqrt{(2N)!}}$ from Eq. \eqref{eq:AEA}. This leads to
	\begin{equation}
		\rho_{0,0}^{\prime\prime} \rho_{2N,2N}^{\prime\prime} - \rho_{2N,0}^{\prime\prime} \rho_{0,2N}^{\prime\prime} =0 - \frac{(2N)!}{16^{N} (N!)^{2}} \rho_{N,N}^{2} < 0,
	\end{equation}
again confirming the nonclassicality of the original state. Note that the above results under both of DM1 and DM2 hold regardless of $\theta$ (quadrature axis), which makes our criteria experimentally favorable. That is, the nonclassicality for an arbitrary FDS state can be verified by observing a single marginal distribution along any directions.
\end{widetext}

\section*{S6. Detection of non-Gaussian states in infinite dimension}
We here consider the detection of nonclassicality for non-Gaussian states in infinite dimension, which has practical relevance to CV quantum information processing. In particular, we examine those non-Gaussian states without squeezing effect in order to demonstrate the merit of our formalism while employing a marginal distribution.

(i) {\bf photon-added coherent state $\hat{a}^{\dag} \ket{\gamma}$}--- We note that a photon-added coherent state do not have squeezing effect for $| \gamma | < 1$ \cite{Agarwal1991}, which restricts the detection of its nonclassicality under a variance test. In contrast, our criteria detect it for all $\gamma$.
We can recast the state as a displaced FDS,
	\begin{align}
		\hat{a}^{\dag} \hat{D} ( \gamma ) \ket{0} & = \hat{D} ( \gamma ) ( \hat{a}^{\dag} + \gamma^{*} ) \ket{0} \nonumber \\
		& = \hat{D} ( \gamma ) ( \ket{1} + \gamma^{*} \ket{0} ),
	\end{align}
where we have used $\hat{D} ( \gamma ) \hat{a}^{\dag} \hat{D}^{\dag} ( \gamma ) = \hat{a}^{\dag} + \gamma^{*}$. As our DM criteria are able to detect every FDS and invariant under displacement operation, we can detect every photon added coherent state using our criteria. Remarkably, our DM methods successfully detect its nonclassicality regardless of quadrature axis as for the case of FDS.

(ii) {\bf photon-added thermal state $\rho = \hat{a}^{\dag} \rho_{\mathrm{th}} ( \bar{n} ) \hat{a}$}--- Its Wigner function is given by
	\begin{align}
		W_{\rho} ( q, p ) & = \frac{2}{\pi} \exp \bigg[ - \frac{2(q^{2}+p^{2})}{1+2n} \bigg] \nonumber \\
		& \times \frac{4(1+n)(q^{2}+p^{2})-(1+2n)}{(1+2n)^{3}},
	\end{align}
with its marginal distribution given by
	\begin{align}
		M_{\rho} ( x ) & = \sqrt{\frac{2}{\pi}} \exp \bigg( - \frac{2q^{2}}{1+2n} \bigg) \nonumber \\
		& \times \frac{n(1+2n)+4(1+n)q^{2}}{(1+2n)^{5/2}}.
	\end{align}
Applying DM method to the marginal distribution and obtaining relevant density matrix elements, we consider $\{ \ket{0}, \ket{4} \}$ subspace for DM1,
	\begin{equation}
		\begin{pmatrix}
			\dfrac{1}{4(1+n)} & - \dfrac{\sqrt{\frac{3}{2}}}{4(1+n)^{3}} \\
			- \dfrac{\sqrt{\frac{3}{2}}}{4(1+n)^{3}} & \dfrac{n^{2}(6+8n+n^{2})}{4(1+n)^{5}}
		\end{pmatrix},
	\end{equation}
and $\{ \ket{0}, \ket{2} \}$ subspace for DM2,
	\begin{equation}
		\begin{pmatrix}
			\dfrac{1}{2\sqrt{1+n}} & \dfrac{2+n}{4\sqrt{2}(1+n)^{3/2}} \\
			\dfrac{2+n}{4\sqrt{2}(1+n)^{3/2}} & \dfrac{3n(4+n)}{16(1+n)^{5/2}}
		\end{pmatrix},
	\end{equation}
then a negative eigenvalue can be found at low mean photon number $n \lesssim 0.40$ and $n \lesssim 0.45$, respectively. For a higher mean photon number, we need to investigate density matrix elements with higher Fock numbers.

(iii) {\bf Dephased odd cat states}---$\rho_{\mathrm{odd}} \sim \ketbra{\gamma}{\gamma} + \ketbra{-\gamma}{-\gamma} - f ( \ketbra{\gamma}{-\gamma} + \ketbra{-\gamma}{\gamma} )$ where $\ket{\gamma}$ is a coherent state with amplitude $\gamma$ and $1-f$ denotes the degree of dephasing. They show no squeezing for all $\gamma > 0$ and $f > 0$. As we obtain the density matrix elements for $\mathcal{D}_{j} [ W_{\rho_{\mathrm{odd}}} ] ( x, y )$ employing the marginal distribution for the rotated quadrature $q_{\theta}$,
	\begin{align}
		M_{\rho_{\mathrm{odd}}} ( q_{\theta} ) = & \sqrt{\frac{2}{\pi}} e^{-2q_{\theta}^{2}-2\gamma^{2}\cos^{2} \theta} \nonumber \\
		\times & \frac{\cosh ( 4 \gamma q_{\theta} \cos \theta ) - f \cos ( 4 \gamma q_{\theta} \sin \theta )}{1 - f e^{-2 \gamma^{2}}},
	\end{align}
we find a non-positive submatrix for $\gamma > 0$ and $f > 0$. More explicitly, for the case of $\mathcal{D}_{1} [ W_{\rho_{\mathrm{odd}}} ] = W_{\rho_{\mathrm{odd}}^{\prime}}$, we find negativity in $\{ \ket{0}, \ket{4} \}$ subspace,
	\begin{align}
		\bra{0} \rho_{\mathrm{odd}}^{\prime} \ket{0} & = \frac{e^{2 \gamma^{2} \sin^{2} \theta} ( e^{\gamma^{2}} - f )^{2}}{(e^{2 \gamma^{2}} - f)^{2}}, \nonumber \\
		\bra{0} \rho_{\mathrm{odd}}^{\prime} \ket{4} & = \bra{4} \rho_{\mathrm{odd}}^{\prime \prime} \ket{0} \nonumber \\
		& = - \frac{\gamma^{4} e^{2\gamma^{2}\sin^{2}\theta}}{4 \sqrt{6} (e^{2\gamma^{2}}-f)^{2}} \{ 3 (e^{\gamma^{2}}+f)^{2} \nonumber \\
		& + 4 (e^{2\gamma^{2}}-f) \cos (2\theta) + (e^{\gamma^{2}}-f)^{2} \cos(4\theta) \}, \nonumber \\
		\bra{4} \rho_{\mathrm{odd}}^{\prime} \ket{4} & = \frac{\gamma^{8} e^{2\gamma^{2}\sin^{2}\theta}}{12(e^{2\gamma^{2}}-f)^{2}} [ 8 e^{2\gamma^{2}} \cos^{8} \theta \nonumber \\
		& + f \{ - e^{\gamma^{2}} \cos^{4} (2\theta) + 8f \sin^{8} \theta \} ],
	\end{align}
particularly its negative determinant
	\begin{align}
		& \bra{0} \rho_{\mathrm{odd}}^{\prime} \ket{0} \bra{4} \rho_{\mathrm{odd}}^{\prime} \ket{4} - |\bra{0} \rho_{\mathrm{odd}}^{\prime \prime} \ket{4}|^{2} \nonumber \\
		& = - \frac{f e^{\gamma^{2}(1+4\sin^{2}\theta)} \gamma^{8}}{12(e^{2\gamma^{2}}-f)^{4}} \{ 9(e^{2\gamma^{2}}+f^{2}) \nonumber \\
		& + 12(e^{2\gamma^{2}}-f^{2})\cos(2\theta) + 4(e^{\gamma^{2}}-f)^{2} \cos(4\theta) \},
	\end{align}
for all $\gamma > 0$ and $f > 0$ independent of the quadrature axis. Note that $9(e^{2\gamma^{2}}+f^{2}) + 12(e^{2\gamma^{2}}-f^{2}) t^{2} + 4(e^{\gamma^{2}}-f)^{2} ( 2t^{2} - 1 )$ is positive for every $t$.


For the case of $\mathcal{D}_{2} [ W_{\rho_{\mathrm{odd}}} ] = W_{\rho_{\mathrm{odd}}^{\prime \prime}}$, we obtain density matrix elements in $\{ \ket{0}, \ket{2} \}$ subspace as
	\begin{align}
		\bra{0} \rho_{\mathrm{odd}}^{\prime \prime} \ket{0} & = \frac{e^{\gamma^{2} \sin^{2} \theta} ( e^{\gamma^{2}} - f )}{e^{2 \gamma^{2}} - f}, \nonumber \\
		\bra{0} \rho_{\mathrm{odd}}^{\prime \prime} \ket{2} & = \bra{2} \rho_{\mathrm{odd}}^{\prime \prime} \ket{0} \nonumber \\
		& = \frac{\gamma^{2} e^{\gamma^{2} \sin^{2} \theta} ( e^{\gamma^{2}} \cos^{2} \theta + f \sin^{2} \theta )}{\sqrt{2} ( e^{2 \gamma^{2}} - f )}, \nonumber \\
		\bra{2} \rho_{\mathrm{odd}}^{\prime \prime} \ket{2} & = \frac{e^{\gamma^{2} \sin^{2} \theta} \gamma^{4} ( e^{\gamma^{2}} \cos^{4} \theta - f \sin^{4} \theta )}{2 (e^{2 \gamma^{2}} - f)},
	\end{align}
with its negative determinant
	\begin{align}
		& \bra{0} \rho_{\mathrm{odd}}^{\prime \prime} \ket{0} \bra{2} \rho_{\mathrm{odd}}^{\prime \prime} \ket{2} - |\bra{0} \rho_{\mathrm{odd}}^{\prime \prime} \ket{2}|^{2} \nonumber \\
		& = - \frac{f e^{\gamma^{2} (1+2\sin^{2}\theta)}\gamma^{4}}{2 (e^{2\gamma^{2}}-f)^{2}},
	\end{align}
for all $\gamma > 0$ and $f > 0$ independent of the quadrature axis.

\bibliographystyle{apsrev}

\begin{thebibliography}{99}
\bibitem{PN} For Gaussian states, our method, if directly applied, works only for the squeezed axes not covering the whole range of quadrature axis. As we show in \cite{supple}, however, a phase-randomization, which does not create nonclassicality, modifies a Gaussian state to a non-Gaussian state for which nonclassicality can be detected regardless of quadrature axis.
\end{thebibliography}

\end{document}